\documentclass[twocolumn]{aastex631}
\usepackage{graphicx} 
\usepackage{xcolor}
\usepackage{textgreek}
\usepackage{amsmath}
\usepackage{amssymb}
\usepackage{hyperref}
\usepackage{float}
\usepackage{wrapfig}
\usepackage{fancyhdr}

\newcommand{\cii}{\ensuremath{{\mathrm{ [C{\normalfont\textsc{ii}}] }}} }
\newcommand{\mum}{$\mu$m}

\newcommand{\ngal}{$n_{\rm Gal}(z)$ }
\newcommand{\ICII}{$I_{[CII]}(z)$ } 

\newcommand{\degsq}{$\mathrm{deg}^2$}

\newcommand{\LxG}{LIM-Gal }
\newcommand{\Var}{\ensuremath{\mathrm{Var}}}

\newcommand\UCLA{Department of Physics and Astronomy, University of California, Los Angleles, CA 90025 , USA}
\newcommand\CIT{California Institute of Technology, 1200 E California Blvd, Pasadena, CA 91125, USA}
\newcommand\JPL{Jet Propulsion Laboratory, Caltech Institute of Technology, Pasadena, CA 91109, USA}
\newcommand\penn{Department of Physics and Astronomy, University of Pennsylvania, Philadelphia, PA 19104, USA}
\newcommand\UIUCA{Department of Astronomy, University of Illinois Urbana-Champaign, Urbana, IL 61801, USA} 
\newcommand\UIUCP{Department of Physics, Grainger College of Engineering, University of Illinois Urbana-Champaign, Urbana, IL 61801, USA} 

\newcommand\UIUCCAPS{Center for Astrophysical Surveys, National Center for Supercomputing Applications, University of Illinois Urbana-Champaign, Urbana, IL 61801, USA}
\newcommand\NCSA{National Center for Supercomputing Applications, University of Illinois Urbana-Champaign, Urbana, IL 61801, USA}
\newcommand\UChicago{Department of Astronomy and Astrophysics, University of Chicago, Chicago, IL 60637, USA}

\newcommand\UA{Department of Astronomy, University of Arizona, Tucson, AZ  85721, USA}
\newcommand\ASU{School of Earth and Space Exploration, Arizona State University, Tempe, AZ 85287, USA}

\begin{document}



\title{Forecasting the cross correlation of Terahertz Intensity Mapper \ensuremath{{\mathrm{ [ C{\normalfont\textsc{ii}}]}}} line intensity maps with Euclid galaxies.}

\author[0009-0001-1134-8729]{Justin S. Bracks}
\affiliation{\UCLA}
\affiliation{\CIT}
\affiliation{\penn}

\author[0000-0003-1859-9640]{Ryan P. Keenan}
\affiliation{Max-Planck-Institut für Astronomie, Königstuhl 17, D-69117 Heidelberg, Germany}

\author[0000-0003-2429-5811]{Shubh Agrawal}
\affiliation{\penn}

\author[0000-0002-3490-146X]{Garrett K. Keating} \affiliation{Center for Astrophysics, Harvard \& Smithsonian, Cambridge, MA, 02067, USA}

\author[0000-0002-4810-666X]{James E. Aguirre}
\affiliation{\penn}

\author[0000-0002-3950-9598]{Adam Lidz}
\affiliation{\penn}

\author[0000-0001-5261-7094]{{Charles M.} Bradford}\affiliation\JPL
\affiliation\CIT
\author[0000-0001-6691-7557]{Brockton Brendal}\affiliation\UIUCP
\author[0000-0001-8217-6832]{Jeffrey P. Filippini}\affiliation\UIUCP
\author[0000-0002-3767-299X]{Jianyang Fu}\affiliation{Department of Physics, Princeton University, Princeton, NJ 08544, USA}
\author[0000-0001-7145-549X]{Karolina Garcia}\affiliation\NCSA
\author[0000-0002-2021-1628]{Christopher Groppi}\affiliation{Department of Physics and Applied Physics, University of Massachusetts, Lowell, Lowell, MA 01854, USA}
\author[0000-0002-8504-7988]{Steve Hailey-Dunsheath}\affiliation\CIT
\author[0000-0001-9122-9668]{Reinier M.J. Janssen}\affiliation\JPL \affiliation\CIT
\author[0009-0006-8665-5754]{Wooseok Kang}\affiliation\UChicago
\author[0009-0004-1270-2373]{Lun-jun Liu}\affiliation\CIT
\author[0000-0003-4063-2646]{Ian Lowe}\affiliation\UA
\author[0000-0003-4629-5759]{Alex Manduca}\affiliation\penn
\author[0000-0002-2367-1080]{Daniel P. Marrone}\affiliation\UA
\author[0000-0001-6397-5516]{Philip Mauskopf}\affiliation\ASU
\author[0000-0001-6439-8140]{Evan C. Mayer}\affiliation\UA
\author[0009-0006-9099-5188]{Sydnee O'Donnell}\affiliation\UIUCP
\author[0009-0001-6336-6793]{Talia Saeid}\affiliation\ASU \affiliation\JPL
\author[0009-0006-8752-1424]{Simon Tartakovsky}\affiliation{Department of Physics, Princeton University, Princeton, NJ 08544, USA}
\author[0000-0002-7058-1221]{Mathilde Van Cuyck}\affiliation\UIUCA
\author[0000-0001-7192-3871]{Joaquin Vieira}\affiliation\UIUCA \affiliation\UIUCP \affiliation\UIUCCAPS
\author[0000-0003-2375-0229]{Jessica A. Zebrowski}\affiliation \UChicago \affiliation{Fermi National Accelerator Laboratory, Batavia, IL, 60510, USA}
\affiliation{Kavli Institute for Cosmological Physics, University of Chicago, Chicago, IL, 60637, USA}

\begin{abstract}


We forecast that the Terahertz Intensity Mapper (TIM) cross-correlated with Euclid’s Fornax deep field (EDF-F), TIM$\times$EDF-F, will detect the \ensuremath{{\mathrm{ [C{\normalfont\textsc{ii}}]}}}-galaxy cross-power spectrum at a median redshift of 1.1 with $\gtrsim 7 \sigma$ confidence. The Poisson component of the cross-power spectrum at 
$0.1 \leq k \leq 10$ hMpc$^{-1}$ (i.e. cross-shot noise) will be detected at $\gtrsim 3 \sigma$ in 4 bins spanning $0.5 < z < 1.7$.  This measurement will constrain the mean \cii specific intensity over half of cosmic history and assess the degree to which Euclid-selected galaxies account for the \cii intensity observed by TIM. We find that TIM can detect the cross-power spectrum across a wide range of \cii intensity models.

\end{abstract}

\section{Introduction}

\begin{figure*}[ht]
\begin{centering}\includegraphics[width=0.98\textwidth]{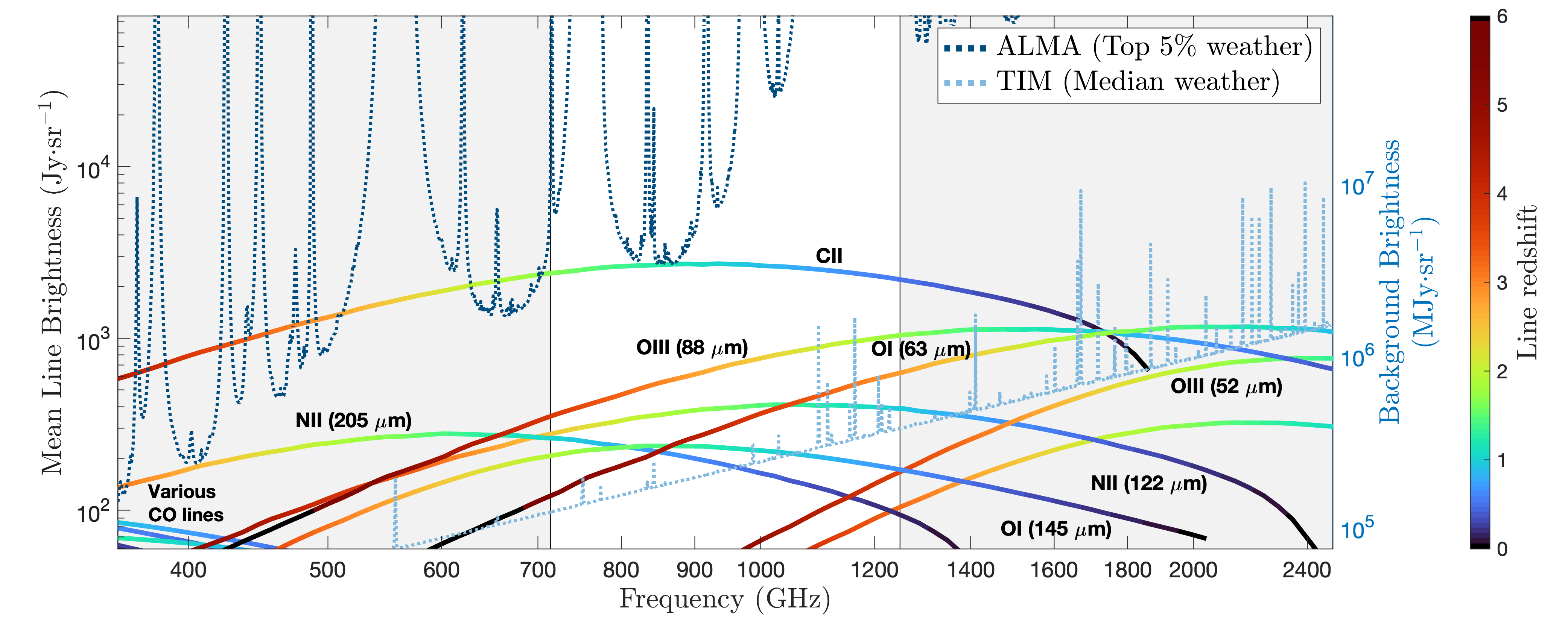}
        \caption{The specific intensity of select FIR line species (left y-axis) versus observed frequency. The dotted lines (using right y-axis) represent the thermal atmospheric background brightness at the ALMA site (dark) and at typical scientific ballooning altitudes (light). The color bar denotes the redshift corresponding to line emission at a given observed frequency. The white region shows TIM's range of operation.
        Predicted line intensities for CO are based on models from \citet{keating20}, calibrated on data from \citet{Kamenetzky16}. Line intensity predictions for \cii, $[\rm NII]_{\rm 122}$, $[\rm NII]_{\rm 205}$, $[\rm OI]_{\rm 63}$, and $[\rm OIII]_{\rm 52}$ are based on empirical models from the SimIM modeling framework \citep{keenan20, keenan22, keating20}. Line intensity predictions for $[\rm OI]_{\rm 145}$ and $[\rm OIII]_{\rm 88}$ are based on models from \citet{bonato19}. The atmospheric brightness is derived using the AM software package \citep{atmospheric_model}.
        }
    \label{fig:kartogram}
\end{centering}
\end{figure*}

Understanding the history of cosmic star formation and its connection to galaxy evolution is one of the most important challenges in modern astrophysics.
The nature of the galaxies responsible for the bulk of star formation has evolved over cosmic time \citep{casey14}. 
Half of the total energy output from the cosmic star formation has been absorbed by interstellar dust and re-emitted in the far-infrared (FIR) \citep{hauser01, lagache05}.
Most star formation has occurred in dust-obscured environments \citep{madau14,zavala+21}, and the most powerful starburst events in cosmic history are effectively invisible at ultraviolet and optical wavelengths \citep{walter12,marrone18}.
The path to understanding galaxy evolution thus necessarily runs through observations in the mid- and far-infrared, and 
requires reliable indicators of obscured star formation. 

The energy output from cosmic star formation is partially reprocessed into dust continuum emission, but also into a variety of bright, unextincted FIR lines. 
These lines provide powerful diagnostics of star formation rates (SFR), supermassive black hole accretion rates, the 
spectrum of ionizing radiation, and the metallicity of the interstellar medium (ISM) \citep{hollenbach97, hollenbach99,Meixner19, schaan&white21Applications, bernal&Kovetz22}. 

The \ensuremath{{\mathrm{ [C{\normalfont\textsc{ii}}]}}}-158 \mum\ line is particularly important, as it is a primary coolant of the ISM \citep{draine11}.  
\cii\ is typically the brightest gas-phase feature emitted by galaxies, accounting for up to 1\% of the total FIR luminosity \citep{draine11}. This makes \cii a promising observational tracer of cosmic star formation. Its brightness and ubiquity within haloes also make it a promising probe for mapping the large-scale structure of the universe \citep{fonseca17,moradinezhaddizgah19,karkare22}.
Careful empirical work has gone into calibrating \cii\ as a tracer of the star-formation rate \citep{delooze+14,herrera-camus+15}, while the advent of ALMA has established the line as a powerful probe of ISM properties at redshifts beyond $z\geq4$ \citep[e.g.,][]{lefevre+20,spilker22,Topping22,Schraerer20}. 
Studies of \cii\ emission are helping to reveal dust-obscured star formation at these high redshifts \citep{loiacono+21, gruppioni+20} and are likely necessary to complement the rest-frame UV view of high-redshift star formation provided by JWST \citep{williams+23,harikane+23, schouws25}. 

However, the Earth's atmosphere is partially opaque and emissive at mid- to far-infrared wavelengths, making observations of \cii\ and other FIR lines at $z\lesssim4$ very difficult from ground-based facilities. Many FIR observations are only possible from stratospheric or space-based platforms. Figure \ref{fig:kartogram} presents the relative line brightnesses for several key emission lines, and contrasts the atmospheric brightness at the ALMA site (considering only the best 5\% of weather days) with that at a scientific balloon's float altitude ($\sim 36$km). Here we see that \cii is the dominant emission line across most of the FIR regime ($< 2$ THz). At wavelengths that target the redshifted \cii line at $z \sim 1$, a ground-based telescope contends with at least an order of magnitude greater background brightness than a stratospheric platform.

Additionally, connecting the physical processes observed in individual galaxies to large-scale structure
will require wide-field mapping in the spatial and spectral dimensions. 
Surveys of molecular and atomic lines have been typically limited to small fields of view and subject to significant sampling uncertainty at cosmological redshifts \citep{keenan20}. 

\begin{table*}[t]
\small 
\label{tab:Parameters}
\center{
\begin{tabular}{l|c|c|l}

\hline
\hline
\multicolumn{4}{l}{\bf Full System} \\
\hline

Temperature & \multicolumn{3}{l}{250mK} \\
Diameter & \multicolumn{3}{l}{2.0~m} \\
Illumination & \multicolumn{3}{l}{cryogenic Lyot stop restricts to $\sim$1.9m diameter}\\
Survey Area & \multicolumn{3}{l}{$1^\circ
\times 1^\circ$} \\

Detector Yield & \multicolumn{3}{l}{85\%} \\

Total Transmission & \multicolumn{3}{l}{8\%} \\

Resolving Power & \multicolumn{3}{l}{$R \approx 250$} \\

\hline
\hline
{\bf Spectrometer } 
& \multicolumn{1}{c | }{Short Wavelength} 
&  \multicolumn{1}{c |}{Long Wavelength} 
& \\
\hline

Detector Array Format 
& \multicolumn{1}{c |}{$51 \times 64$} 
& \multicolumn{1}{c |}{$51 \times 64$} 
& spatial $\times$ spectral \\

Wavelength range & 240 -- 317 & 317 -- 420 & $\mu$ m \\
$\Delta \nu$ & 4.3 & 3.3 & GHz \\
Beam FWHM & 37 & 48 & arcsec \\

NEI (median)& 124.5 & 64.2 & $10^7$ Jy/sr $\sqrt{s}$  \\

\cii\ redshift range & 0.52--1.0 & 1.0--1.67 & \\
\cii\ cosmic epoch & 5.9--8.5 & 3.9--5.9 & Gyr of cosmic age \\
\hline
\hline
\end{tabular}
\caption{Design characteristics for TIM. The upper table lists characteristics of the full system or that are shared by both detector modules. The lower table reports properties of the spectrometer for the short and long wavelength modules.}

}

\end{table*}

Line Intensity Mapping (LIM) is an emerging approach for capturing the 3D structure of the universe via large-scale, volumetric measurements \citep[for a review see][]{changLidz26}. The LIM technique measures the amplitude of intensity fluctuations as a function of spatial scale, capturing the statistical characteristics of the emitting sources. 
\cite{keating16, keating20} report initial shot regime detections of CO signal ($k\gtrsim1 $ $h \, \mathrm{Mpc}^{-1}$). There have also been detections using the hydrogen 21 cm line in cross-correlation with traditional galaxy surveys \citep{chang2010, Amiri23, Cunnington23, masui2013, switzer13}.

It is currently challenging for LIM surveys to achieve robust auto-power spectrum measurements, especially since the field is relatively young. A range of systematic challenges need to be overcome \citep{bernal&Kovetz22}. For instance, the IR continuum is expected to be at least 2 orders of magnitude brighter than \cii line emission. Additionally, foreground emitters and line interlopers further confuse potential detections. 

 Fortunately, LIM lends itself very well to cross-correlation between pairs of surveys that cover overlapping redshift ranges with different tracers. This can serve as a means of alleviating or circumventing some of the most insidious challenges. Cross-correlation avoids an average bias from unshared foregrounds and line-interlopers, where emission from a source other than the target line is redshifted into the target spectral band. 

Cross-correlations of line intensity maps with the matter density field, as traced by spectroscopic galaxy surveys (LIM-Gal), will be a particularly potent approach for line intensity mappers targeting low and modest redshift ranges ($z \lesssim  2$) where galaxy surveys can measure relatively large proportions of the galaxy population. The signal-to-noise ratio (SNR) of \LxG measurements is driven, in part, by the number density, \ngal of galaxies in the galaxy survey at the requisite redshifts. Due to their intrinsic flux limits, contemporary galaxy surveys suffer a decline in \ngal beyond $z\approx1$ \citep{momcheva163dHST}. Cross-correlation with a LIM survey across this drop-off in sensitivity can inform how much of the galaxy luminosity function is being missed as a function of redshift. In this regime, an infrared LIM survey in cross-correlation with a spectroscopic galaxy survey wields considerable discovery potential, while also providing robust cross-checks from a tried-and-true method. Further, the comparably high SNR that galaxy surveys bring to the table significantly boosts the SNR in the cross-power spectrum relative to the line's auto spectrum. This approach has already served as a critical proof of concept in 21 cm cosmology \citep{pen2009, chang2010, masui2013, anderson2018, wolz2022, cunnington2023h, switzer13} and been explored for a range of early submillimeter and FIR LIM measurements \citep{Pullen13, Pullen18,  Yang+19, keenan22}.

In this paper, we emphasize the capabilities of the forthcoming Terahertz Intensity Mapper (TIM), a NASA-funded project to carry out a \cii LIM survey from a stratospheric balloon, specifically in cross-correlation with the \textit{Euclid} space telescope's deep field, Fornax.

The paper is organized as follows. 
Section \ref{sec:TIM} presents the as-built, or as-designed characteristics and limitations of the TIM.  Section \ref{sec:Formalism} describes the forecasting formalism for the cross-correlation measurement between a \cii\ LIM experiment and a spectroscopic galaxy survey. Then, we detail any assumptions or approximations that we adopt in our fiducial forecast. 
Section \ref{sec: Results} presents the salient outcomes of our investigation, and reports the implications of our cross-correlation forecasts for constraining the \cii history.
We conclude in section \ref{sec:Conclusion}, and briefly discuss a potential successor experiment to TIM. Finally, we include an appendix, \ref{sec:GalSurvs} where we motivate our decision to use EDF-F as our primary cross-correlator over comparable surveys.

In the following, we assume the Planck 2018 cosmological parameters \citep{Planck18}. In all calculations requiring a star formation rate density (SFRD), we assume Kroupa initial mass functions, converting between IMFs using table 1 from \citet{Kennicutt12} where necessary. 

\section{TIM}
\label{sec:TIM}

A number of \cii line intensity mapping instruments will become operational in the coming years (e.g TIME, CCAT, EXCLAIM) \citep{vieira20, Crites22, Choi23CCAT, Essinger_Hileman20Exclaim}, and numerous second-generation instruments exploiting advances in submillimeter/terahertz instrumentation are also under consideration. At the same time, densely sampled, wide field spectroscopic surveys using space-based grism spectroscopy are now greatly expanding the footprints of galaxy surveys suitable for cross-correlation measurements \citep{euclid24, pozetti16, mowla3dDASH, momcheva163dHST}. Here we overview one such LIM instrument suitable for \cii LIM at $z\sim1$ -- TIM.

The upcoming Terahertz Intensity Mapper experiment \citep{vieira20,marrone2022terahertz} is a balloon-borne far infrared telescope to be flown from Antarctica in the austral summer of 2027-2028. TIM is designed to map \cii at the peak and drop off of cosmic star formation, $0.52 < z < 1.67$ ($240 \mu m \leq \lambda \leq 420 \mu m$). As shown in figure \ref{fig:kartogram}, \cii is the brightest FIR line in the wavelength range we target, reducing contamination from interloper lines. \cii is robust against dust extinction. These qualities make \cii an ideal diagnostic line for characterizing the history and nature of otherwise dust-obscured star forming galaxies \citep{lagache17DSFG_LIM,gruppioni20alpineDSFG}. As shown in figure \ref{fig:kartogram}, the atmosphere is largely opaque to terahertz photons. TIM must make its observations from above the majority of the atmosphere, driving the need for a balloon platform.

\begin{figure}[t!]
\begin{centering}\includegraphics[width=0.48\textwidth]{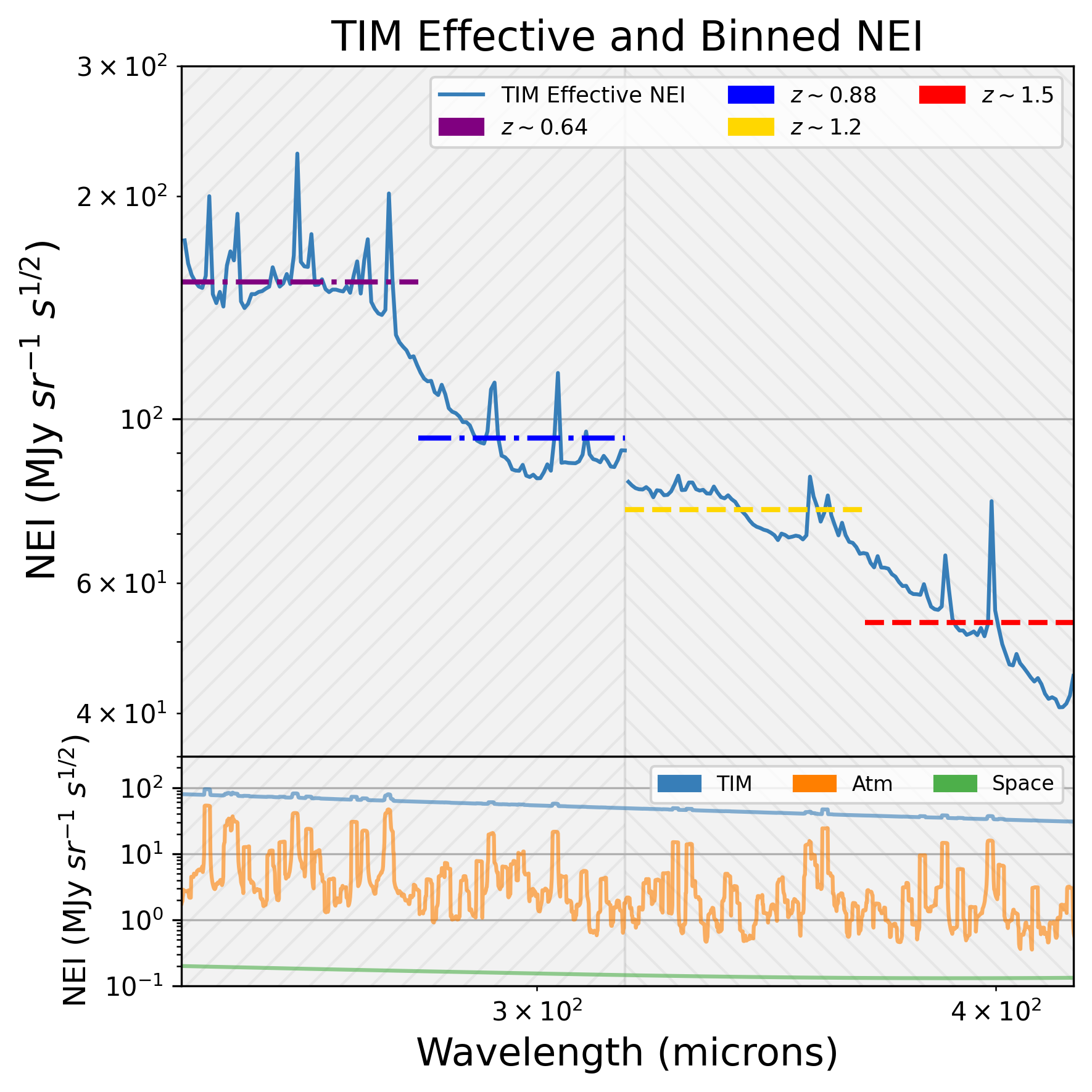}
        \caption{\textbf{Top} Nominal NEI for the TIM detector focal planes. We present the averaged NEI for TIM's 4 redshift bins separately in purple, blue, yellow and red, depicting increasing redshift. Gray-filled areas denote the wavelength bands for TIM's SWA (upward hashed) and LWA (downward hashed). TIM's band-averaged NEI is shown by dashed (SWA) and dash-dotted (LWA) horizontal lines. The NEI for the current TIM instrument is dominated by the Poisson noise imparted by the warm primary mirror. \textbf{Bottom} We include equivalent curves for both a stratospheric atmosphere- (orange) and zodiacal-background-light (green) limited version of the TIM optics.
        }
    \label{fig:NEIs}
\end{centering}
\end{figure}

TIM uses cryogenically cooled kinetic inductance detectors (KIDs). KIDs are a promising approach to large-format far-IR detector arrays in which each detector pixel is a lithographically patterned superconducting microresonator (\citealt{zmuidzinas12} for a review). TIM will employ two optically separated detector modules, the short wavelength array (SWA) (240 \mum - 317 \mum), and the long wavelength array (LWA) (317 microns - 420 microns). A module is comprised of four subarrays each housing roughly 1,000 KIDs. The bandwidths of SWA and LWA are represented in Figure \ref{fig:NEIs} by the upward and downward grey hashed regions, respectively. In the following forecasts we further split the TIM redshift range at the centers of both the SWA and LWA modules to create four TIM redshift bins centered at approximately $z = {0.64, 0.9, 1.2, 1.5}$ The average total noise equivalent intensity (NEI) seen by the TIM detectors in each of these redshift bins is shown in figure \ref{fig:NEIs}. The TIM KIDs noise characteristics are subdominant to the contribution from the experiment's ambient-temperature primary mirror, which is the primary source of system noise. For comparison, we have included the NEI profiles for an atmosphere-limited version of TIM, as well as a background-limited (zodiacal light primarily) version that could be achieved by a space mission.

\section{Cross-spectrum Formalism}
\label{sec:Formalism}
In a cross-spectrum, we consider the correlation between the intensity fields of two tracers. In this section, we review a formalism for cross-correlating two independent tracer phenomena. We begin by adopting the cross-correlation formalism put forth in \cite{Schaan&white21theory} for the correlation of two line emission intensity fields. We then specify this formalism toward the cross-correlation between one emission line and the galaxy overdensity field.

\subsection{The Line-Line Formalism}
Following \citet{Schaan&white21theory}, the redshift- and scale-dependent cross-power spectrum of two surveys can be expressed as the summed contributions from the 2-halo, 1-halo, and shot-power terms, which respectively capture the contributions from the large-scale distribution of matter, the distribution of galaxies within their dark matter halos, and a Poisson term due to the discrete locations of individual emitters:
\begin{equation}
P_{i\times j}(k,z) = P_1^{\rm Halo}(k,z) + P_2^{\rm Halo}(k,z) + P_{\rm shot}^\times(z)
\label{eq:crossPowerFormal}
\end{equation}

The two-halo term can be defined as:
\begin{equation}
P_2^{\rm Halo} = b_i(z) b_j(z) \langle I_i(z) \rangle  \langle I_j(z) \rangle P_{M}(k, z)
\label{eq:two-halo}
\end{equation}
$b_{i}$ and $b_{j}$ are the linear biases of tracers $i$ and $j$ relative to the matter over-density field with power spectrum $P_M$. The average specific intensity of emission i at redshift z, $\langle I_{i}(z) \rangle$ is given by
\begin{equation}
    \langle I_{i} \rangle = \frac{c} {4 \pi H(z) \nu_{i}} \int \phi(L_i)L_i dL_i
    \label{Intensity}
\end{equation}
where $\phi$ is the luminosity function of line $i$. 

The 1-halo term  can be expressed as:
\begin{equation}
P_1^{\rm Halo} = \frac{U^2_\times(k)}{n_{h,\times}} \langle I_i(z)\rangle \langle I_j(z)\rangle
\label{eq:1-halo}
\end{equation}
where $U_\times$ and $n_{h,\times}$ describe the tracer weighted profile and number
density of dark matter halos. $P_1^{\rm Halo}$ is a second-order term with considerable complexity from a modeling standpoint. We omit this term from our fiducial forecast, which modestly suppresses the forecast power at intermediate scales ($k \approx 1$ hMpc$^{-1}$) in comparison to similarly motivated models which include the 1-halo term (perhaps most conveniently illustrated by comparing the ``Keenan+22'' and ``This Work'' lines in figure \ref{fig:constrainingPower}).

Finally, the cross-shot power spectrum can be calculated as 
    \begin{equation}
    P_{\rm shot}^\times = \left(\frac{c}{4\pi H(z)} \right)^2 \frac{1}{\nu_i\nu_j} \int \int dL_i dL_j \Phi(L_i,L_j) L_i L_j
        \label{eq: crossShotFormal}
    \end{equation}
Where $\Phi$ is the joint distribution function describing the number of galaxies per unit volume of luminosities $L_i$ and $L_j$ in the two tracer lines.

\subsection{Modeling the \cii $\times$ Galaxy Overdensity Field Cross-Power Spectrum}
\label{crossCorrChoices}
The cross-correlation between a line intensity map and a galaxy survey can be thought of as a special case of the line-line cross-power spectrum (Eq \ref{eq:crossPowerFormal}), where $I_{\rm Gal}(z)$ is taken to be $1+\delta_{\rm Gal}$ and $\delta_{\rm Gal}$ is the overdensity field of the galaxies in the survey. By construction, the `mean intensity' is unity and unitless; $\langle I_{\rm Gal}\rangle = 1$. The cross-shot noise component arises from the Fourier transform of the self-correlation term, i.e. contributions to $ \langle I_i(\bf{x_1}) \delta_{\rm Gal}(\bf{x_2}) \rangle$ at a single point, $\bf{x_1} = \bf{x_2}$. This reflects the contribution of galaxies in the traditional survey to the specific intensity in the LIM observations. In Fourier space this gives
\begin{equation}
    P_{\rm shot}^\times = f_s \langle I_i \rangle/n_{\rm Gal}
    \label{eq:crossshot}
\end{equation} 
where $n_{\rm Gal}(z)$ is the number density of galaxies in the galaxy catalog and $f_s$ is the fraction of $\langle I_i  \rangle$ that can be attributed to galaxies in the galaxy catalog. 

To obtain a plausible model value for $f_s$, we compare the SFRD in the ASTRODEEP catalogue as presented in \cite{Merlin_2021} to the SFRD curve in \cite{madau14}, resulting in $f_s \sim 0.9$.  

Because both the ASTRODEEP catalog and the EDF-F catalog have similar limiting magnitude and spectral resolution (see table \ref{tab: galaxy_surveys}), we can approximate their $f_s$ values as equal.


Here we are implicitly assuming that $L_{\cii} \propto \rm SFR$ for all systems (similar to \cite{yang+22} and related models). In this case, the fraction of the total \cii intensity produced by the galaxies matches the fraction of the SFRD they produce - modulo scatter in the $L_{\cii}-\rm SFR$ relation. The effect of this scatter is averaged down in a large sample and is negligible.

Continuing with the assumption that $\langle I_{\cii} \rangle$ closely traces star formation, we parameterize the integral in equation \ref{Intensity} as 
\begin{equation}
\int dL_\cii \phi(L_\cii) L_\cii = L_0 \dot{\rho}_*(z)
\end{equation}
Here $\dot\rho_{\star}$ is the SFRD, and $L_0$ is the line emission luminosity from a galaxy with a star formation rate of $1 M_\odot /\mathrm{yr}$ - in other words, the line luminosity per unit SFR. $\dot\rho_\star(z)$ is assumed to follow the parameterized history from \citet{MD_2014}. To determine $L_0^{\cii}$, we assume a simple ratio-metric power law relationship between the average \cii luminosity and the overall IR luminosity, choosing to employ the full galaxy population value presented in Table 3 of \citet{De_Looze_2014} (slope of 1.01). 

The matter power spectrum, $P_M(k,z)$ is approximated by the linear theory prediction, which should be adequate on the spatial scales where the clustering signal dominates over the cross-shot noise contribution $k \lesssim 1 \, h \,\mathrm{Mpc}^{-1}$. We compute $P_M(k,z)$ using
\textit{CAMB} \citep{CAMBL}.
\begin{figure}[b]
\begin{centering}\includegraphics[width=0.48\textwidth]{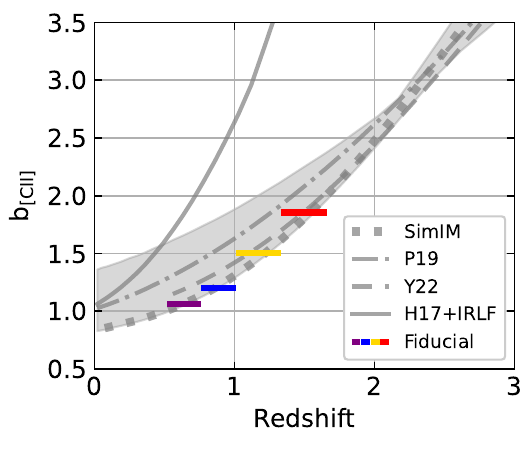}
        \caption{Redshift-dependent linear clustering bias models using separate model classes. The colored bars are the fiducial biases from this work, with the individual colors bearing the same meaning as in figure \ref{fig:NEIs}.
Grey fill region display forecasts using similar models, effectively serving as a literature-based uncertainty on our assumed biases. Dot-dashed gray
line: \cite{padmanabhan+19}, dashed gray line: \cite{yang+22}, dotted grey line: \cite{keenan20}. 
        }
    \label{fig:biases}
\end{centering}
\end{figure}

To develop a quantitative estimate of the bias values for the galaxy and \cii surveys considered in this investigation, we solve the auto-power spectrum ($P_{i\times i} \equiv P_\cii$) of equation \ref{eq:crossPowerFormal} for the $b_i$ parameter in the two-halo term (see eq. \ref{eq:two-halo}).
We then obtain $P_{\cii}$ and $\langle I_{\cii}(z) \rangle$ using SimIM \citep{keenan20}. From this we attain $b_\cii$. For the central redshift of each of TIM's four redshift bins; this yields $b_\cii(z \approx 0.65, 0.89, 1.18, 1.5) \approx 1.22, 1.35, 1.6, 1.87$, which agrees with the $b_{\rm Gal}$ values determined by \cite{Durkalec15} (under the assumption that linear galaxy bias traces star formation) and \cite{Jullo12}. Since we are now assuming that both $b_\cii$ and $b_{\rm Gal}$ closely trace the SFR, as an approximation in our model we set $b_{\rm Gal} = b_{\cii}$. In practice, $b_{\rm Gal}$ can be determined using the galaxy auto-power spectrum as well. Figure \ref{fig:biases} shows our calculated value for $b_{\cii}(z)$ and compared with the analogous $b_{\cii}(z)$ values found by applying the same treatment to other \ICII models in the literature. We see that our values agree well with other models who derive their \ICII history from a \cii- SFR relation such as \cite{yang+22}.

Neglecting the 1-halo term as discussed above, and applying equations \ref{eq:two-halo} and \ref{eq:crossshot}  specifically to a \cii LIM survey in cross-correlation with a spectroscopic galaxy survey as discussed in the introduction, Equation \ref{eq:crossPowerFormal} becomes
    \begin{equation}
    \label{eq:CIIxGal}
    P_{\cii \times \rm Gal} \equiv P_\times  =  b_\cii b_{\rm Gal} \langle I_{\cii} \rangle  P_{M} + f_s \langle I_{\cii}\rangle/n_{\rm Gal}
    \end{equation}
with $z$ and $k$ dependencies suppressed for clarity.

\subsection{Cross spectral variance}
Following \cite{VN15}, the variance attributed to a \LxG cross-power spectrum measurement can be written as:
    \begin{equation}
    \Var[P_\times] = \frac{1}{2N_{\rm mode}}  \left[P^2_\times + P_{\rm Gal} \left(P_{\cii} + \frac{P_N}{\mathcal{B}}\right)\right]
    \label{variance}
    \end{equation}
where $\mathcal{B}$ is the instrument beam effect imposed on the \cii signal. The beam term is discussed in the following sub-section, while the other terms (and special considerations) contributing to the cross-variance are discussed in the following bullets:
\begin{itemize} 
    \item $P_\times$, $P_{\rm Gal}$, and $P_\cii$ (the power spectral terms) account for sample variance in the clustering regime of the cross-power, galaxy auto power, and \cii auto power, respectively, and in most cases (certainly true for TIM) are subdominant to the system noise terms. We have suppressed the scale and redshift dependence of $P_\times$, $P_\cii$, and $P_{\rm Gal}$ for readability. The auto power spectra, $P_\cii$ and $P_{\rm Gal}$ can be expressed using Eq \ref{eq:crossPowerFormal}, except we consider cross correlating each survey with itself such that $P_\cii = b_\cii^2 I_\cii^2 P_M + P_{\rm shot}^{\cii}$ and $P_{\rm Gal} = b_{\rm Gal}^2 P_M + P_{\rm shot}^{\rm Gal}$. We compute $P_{\rm shot}^{\cii}$  using a semi-empirical model for the star formation rate distribution function \citep{behroozi19} and convert this to a [CII] luminosity function using the SimIM modeling framework (Keenan et al. in prep), while
    $P_{\rm shot}^{\rm Gal} = 1/n_{\rm Gal}$.
    
    \item $P_{N}$, the instrument noise power contribution from the \cii survey can be expressed as $P_N = \sigma_P^2 V_{\rm Vox}$. Where $V_{\rm Vox}$ is the comoving volume of an instrument voxel defined by the beam size (sky-transverse direction) and spectral resolution (line-of-sight direction). 
    Here $\sigma_P$ is the contribution from instrumental noise, which is proportional to the overall noise equivalent intensity 
    (NEI) of the experiment in that band. For a survey that reaches a uniform depth across the whole map, $\sigma_p$ can be approximated as:
    \begin{equation}
        \sigma_P \approx NEI \sqrt{\frac{N_{\rm pix}} {N_{\rm det} t_{S}}}
        \label{eq: sigmaP}
    \end{equation}
    Here $t_S$ is the total integration time of the survey (assumed to be $\sim 200$ hrs for TIM), $N_{\rm pix}$ is the number of (2D) elements into which the map is pixelized, $N_{\rm det}$ is the number of detectors observing the same spectral range (see table \ref{tab:Parameters}), or spatially independent spectra observed simultaneously by the instrument.

    \item $N_{\rm mode}$ refers to the 
     number of times an individual mode uniquely occurs in the spherically averaged volume captured by a survey. We discuss some nuances our approach imposes on the number of measurable samples per mode in the following two subsections. 
     

    \end{itemize}

\subsubsection{Beam and Spectral Attenuation}
\label{sec: windowing}
    
An experiment's finite spectral and spatial resolutions impose limitations on the minimum measurable spatial extent in the line-of-sight and transverse directions, respectively. The wave-numbers corresponding to the physical scales represented by the spectral and spatial resolution elements can be calculated as $\sigma_{\parallel}(z) = L_\parallel(z)/\sqrt{8 \ln{(2)}}$ and $\sigma_{\perp}(z) = L_\perp(z)/\sqrt{8 \ln{(2)}}$, where $L_\parallel$ is the comoving distance spanned by the detector spectral resolution (full width at half maximum [FWHM]), $d\nu$ at a redshift $z$, and $L_\perp$ is the comoving distance spanned by the telescope beam FWHM at a specific redshift. As the measured scale approaches the resolution limit, the fidelity of the measurement suffers.
The observed intensity fluctuations are attenuated by the finite angular and spectral resolutions of the instrument. 
We approximate the beam response as a Gaussian, which disperses the sky intensity in the perpendicular direction according to:
\begin{equation}
b(x,y) = \frac{1}{2\pi\sigma_{\perp}^2} \exp\Big[-\frac{1}{2}\frac{x^2+y^2}{\sigma_\perp^2}\Big],
\end{equation} 
For simplicity, we also assume a Gaussian spectral response;
\begin{equation} 
b(z) = \frac{1}{\sqrt{2\pi\sigma_{\parallel}^2}} \exp\Big[-\frac{1}{2}\frac{z^2}{\sigma_\parallel^2}\Big].
\end{equation}

We can combine the spatial and spectral resolution contributions into a three-dimensional attenuation function $B(x,y,z) = b(x,y) b(z)$.
The Fourier transform $\mathcal{F}[B(x,y,z)]$ gives a wave number dependent transfer function describing the finite resolution effects, $\overline{B}(k_x, k_y, k_z)$:
\begin{equation} 
\label{bBar}
 \overline{B} (\textbf{k}) =  \exp\Big[-\frac{1}{2}(k_x^2+k_y^2)\sigma_\perp^2-\frac{1}{2}k_z^2\sigma_\parallel^2\Big]
\end{equation} 
where $k_x$, $k_y$ and $k_z$ are components of an individual mode, $\textbf{k}$. Similarly, the survey geometry and spectral bandwidth drive the maximum measurable spatial extents, which define the total survey \textit{volume}, and therefore the minimum k modes. We take the conservative approach of removing any mode which has $k_{i} \approx 0$. That is, we do not count a mode if any of its directional components is 0. This has the effect of depressing $N_{\rm mode}$ where $k < 0.1 \, h$  Mpc$^{-1}$.


%
%

In the case of \LxG where the positional accuracy of the galaxy survey far outstrips the LIM survey, we can treat the galaxy survey's $x$ and $y$ attenuation contributions as delta functions. That is, the only contribution to the attenuation function coming from the galaxy survey stems from the survey's spectral uncertainty. 

For our analysis, we define a linear average of $\overline{B}$ coefficients with a bin of width $d\ln(k)$:
\begin{equation}
\mathcal{B} =  \frac{1}{N_k}\Sigma_{k - \delta k}^{k + \delta k} \overline{B}
\end{equation}

%
%
%

\subsubsection{Survey Area and $N_{\rm mode}$}

In general, there is a practical trade-off between sky coverage and survey depth. 
Increasing the survey area augments the number of samples per mode. This could decrease variance without the need to make any hardware changes. However, increasing survey area further divides the survey time, resulting in less integration time per voxel and leading to a corresponding increase in $P_N$.

In the case of the auto-power, the optimal survey area is simply chosen such that $P_N \approx P_\cii(k)$. A larger area will inflict a considerable noise penalty. However, this is not necessarily the case in the cross-spectrum. For our purposes, we discuss this in the context of a \cii $\times$ Galaxy Survey cross-spectrum, but the following is generally true for LIM cross-spectra: 
In the limit that the LIM measurement variance is dominated by instrumental noise, i.e. $P_N \gg P_\cii$ and $P_N P_{\rm Gal} \gg P^2_\times$, Eq \ref{variance} becomes:
\begin{equation}
    \Var[P_\times] = \frac{1}{2N_{\rm mode}} \left(\frac{P_N}{\mathcal{B}} P_{\rm gal}\right) 
    \label{noiseLimited}
\end{equation}
\begin{figure*}[!t]
\begin{centering}\includegraphics[width=0.98\textwidth]{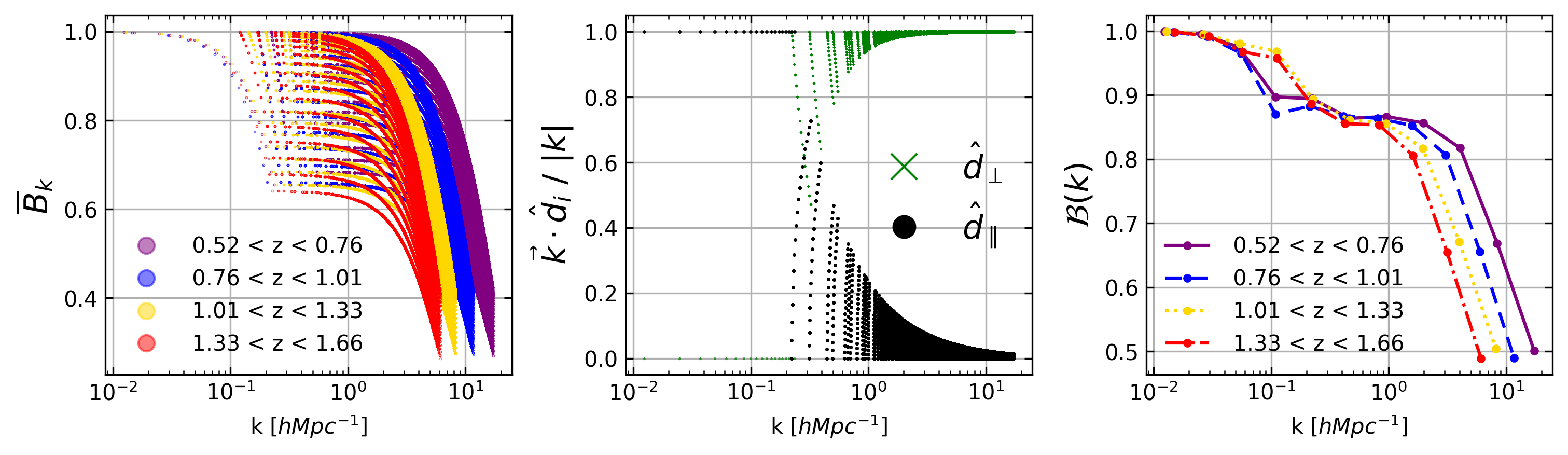}
        \caption{\textit{\textbf{left}}: $\overline{B}(k)$, the transfer coefficient for each individual k mode measured by TIM in each of its 4 redshift bins. \textit{\textbf{center}:} Projection of each mode $\textbf{k}$ in the transverse ($k_\perp$) and line-of-sight ($k_\parallel$) directions for TIM's lowest redshift ($0.52 < z < 0.77$) bin. \textit{\textbf{right}:} The linear average of $\overline{B}(k)$ within logarithmically spaced, spherical shells in $k$-space denoted $\mathcal{B}(k)$.}
    \label{fig:window}
\end{centering}
\end{figure*}
Because both $N_{\rm mode}$ and $P_N$ scale linearly with the survey area, this limit yields the surprising result that (for a particular $k$ mode) there is no optimal survey area. That is, for the cross-spectrum, as the survey area grows, the map noise increases owing to the decreasing observing time per voxel, but this is exactly compensated by the gain in the number of modes sampled. Note that this supposes the galaxy survey spans the entire enlarged \cii survey area at adequate depth. It also assumes that the modes of interest are well-sampled, i.e., their wavelengths are small compared to the survey dimensions and captured given the angular/spectral resolution of the \cii experiment.

%
%
%
%

\subsection{Retrieving $\langle I_{\cii} \rangle$ and SFRD(z)}

By isolating and dividing out $\langle I_{\cii} \rangle$ from equation \ref{eq:CIIxGal} we obtain: 
\begin{equation}
\langle I_{\cii} \rangle = \frac{P_\times} {b_\cii b_{\rm Gal} P_M + f_s/n_{\rm Gal}}. 
\end{equation}

If the terms in the denominator are known, an estimate of \ICII can then be derived from the inverse-variance-weighted average over all $k$. Given the modest SNR of TIM, $P_M$ and $b_{\rm Gal}$ can be treated as known constants derived from the standard cosmological model and the galaxy survey respectively. The remaining terms, $b_\cii$ and $f_s$ must be approximated, and contribute an additional systematic uncertainty to the derivation of \ICII. The range of empirical models in Figure~\ref{fig:biases} (gray filled region) suggests this uncertainty is around $\pm20$\% for $b_\cii$. 
Because the \textit{Euclid} deep fields (as TIM's primary cross-correlator) will probe significantly below the knee of the H$\alpha$ luminosity function, we expect the galaxy catalog to account for the majority of the cosmic star formation rate density and \ICII. This suggests $f_s\gg50$\%, with correspondingly small uncertainties. Estimates of $f_s$ will improve once the depth and completeness of the \textit{Euclid} deep fields is better characterized \citep{euclidQ1Spec}.

\section{Results}
\label{sec: Results}

\begin{figure*}[t]
\begin{centering}\includegraphics[width=0.98\textwidth]{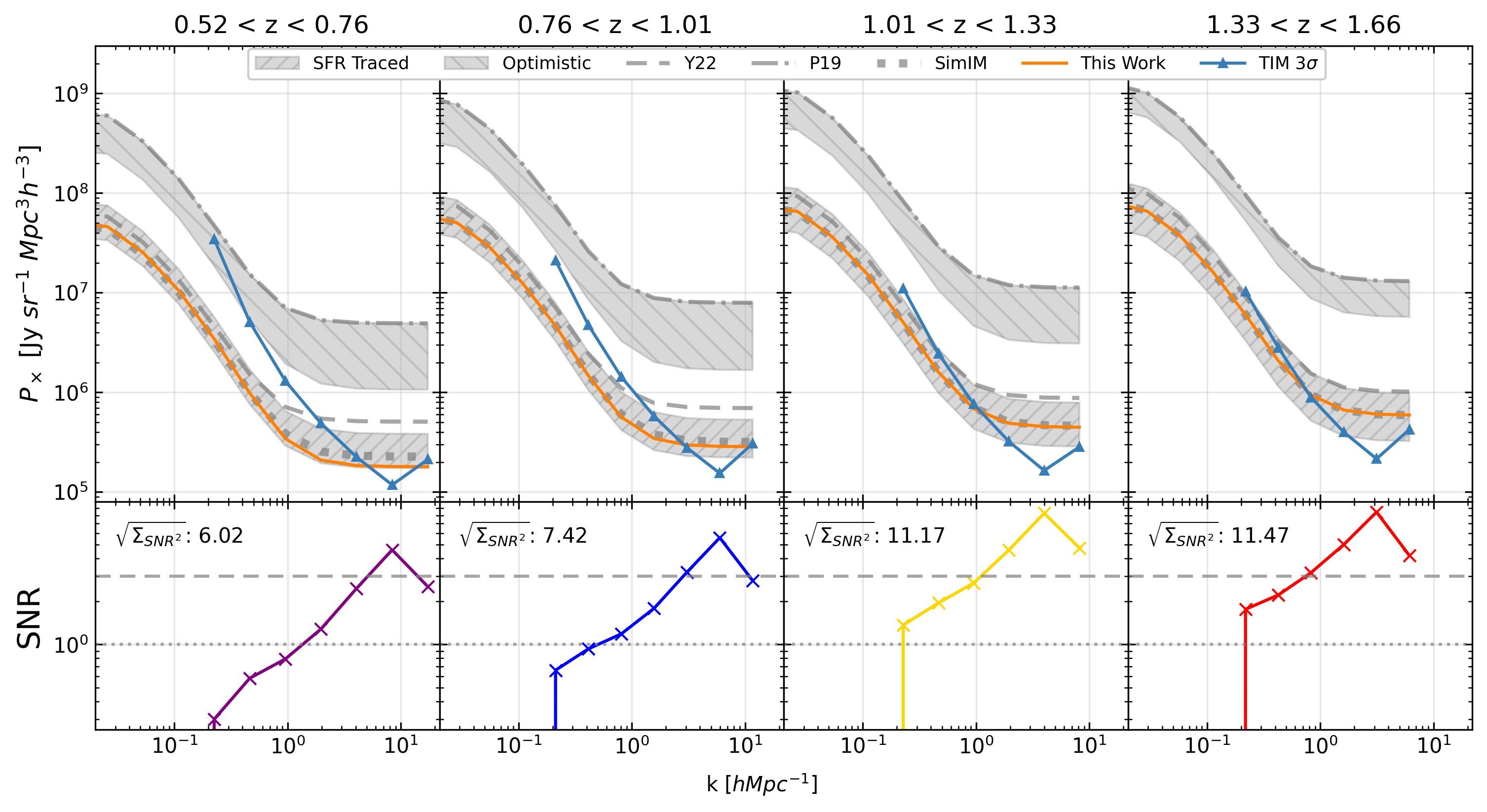}
        \caption{
        \textbf{Top}: Forecast TIM $P_{\cii\times \rm EDF-F}$ sensitivity curves in 4 redshift bins. Upward-facing arrows represent the $1\sigma$ sensitivity; blue solid line represents TIM as designed with a 1 $\mathrm{deg}^2$ field; orange solid line the $P_{\cii\times \rm EDF-F}$ forecast from this work. Grey fill regions display analogous forecasts using parameters from alternative models in the current literature.
        Dot-dashed gray line: \cite{padmanabhan+19}, Dashed gray line: \cite{yang+22}, dotted grey line: \cite{keenan20}.
        \textbf{Bottom}: SNR per $k$ bin, with total expected per-redshift-bin SNR presented in the upper left of each SNR subfigure. Dotted and dashed grey lines present 1$\sigma$ and 3$\sigma$ thresholds respectively.
        }
    \label{fig:constrainingPower}
\end{centering}
\end{figure*}

In this section we explore spectral and spatial resolution attenuation effects specific to TIM, discuss TIM's ability to sample spatial scales with respect to $N_{\rm mode}(k)$ and finally present TIM's ability to constrain models of $\langle I_{\cii}\rangle$ and the cosmic star formation history.

The three panels in figure \ref{fig:window} show the scale-dependent behavior of the beam window function (as formally introduced in Section \ref{sec: windowing}) in the four TIM redshift bins. In the \textbf{leftmost panel} we present the $\overline{B}$ coefficients for each unique mode sampled by the baseline TIM experiment in cross-correlation with projected \textit{Euclid} capabilities. Here we see that the bulk of the available $k$ mode samples are concentrated toward higher $k$, or smaller scales. Then, as the physical scales approach the comoving distance subtended by the telescope beam, information is lost and $\overline{B}$ falls off steeply.

\textbf{The center panel} of figure \ref{fig:window} shows the projection of each mode vector onto the line-of-sight direction. For clarity, this is only shown for TIM's lowest redshift bin ($0.52 < z < 0.77$) but the other redshift bins behave similarly. This is shown to illustrate that modes below $k \sim 0.2$ $h \, \mathrm{Mpc}^{-1}$ are \textit{only} captured in the line-of-sight ($k_\parallel$) direction. This is due to the fact that TIM's line-of-sight extent represents a much larger comoving scale than the largest scales probed in the transverse directions. Similarly, the physical scales subtended by the spectral resolutions of both TIM's long- and short-wavelength modules translate to considerably greater physical extents than TIM's spatial resolution, again meaning that each of TIM's voxels are much longer in the spectral/line-of-sight direction than they are tall or wide in the spatial/transverse directions. 


As we approach higher $k$'s the contribution from the transverse ($k_\perp$) component increases, dominating by $k \sim 1~ h$Mpc$^{-1}$. This transition from line-of-sight to transverse dominated regimes is reflected in the double-hump profile in \textbf{the rightmost panel} of figure \ref{fig:window}. Here the average beam attenuation coefficient of samples in logarithmically spaced $k$ bins tracks the few line-of-sight modes at low $k$, then transitions to tracking the lower coefficient, but much more densely sampled transverse-direction-dominated modes at high k. This affects the scale-dependent sensitivity of the TIM \cii survey.

\subsection{Sensitivity Forecast for the TIM \cii Survey in Cross-Correlation with EDF-F} 

\begin{figure*}[t]
\begin{centering}\includegraphics[width=0.98\textwidth]{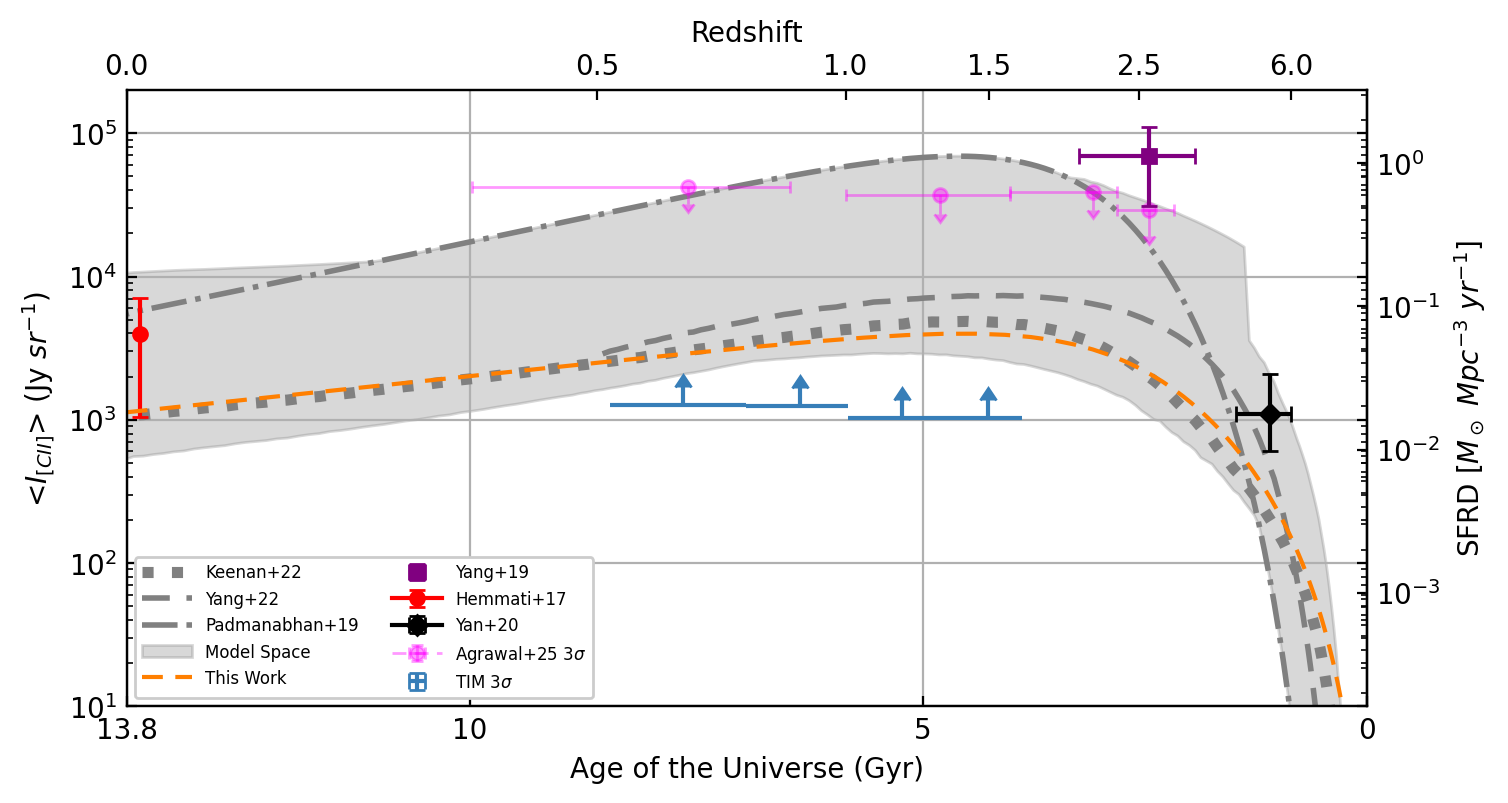}
        \caption { Forecast constraints on the mean specific intensity (left y axis) history of \cii, and correspondingly, the cosmic SFRD (right y axis) as a function of cosmic age (lower x axis) or redshift (upper x axis). The colored horizontal bars with upward-facing triangles represent the $3\sigma$ detection threshold forecasts for each of TIM's 4 redshift bins. The pink bars with downward-facing arrows represent the $3\sigma$ upper limit found in \citep{Agrawal+25}; the bar widths represent the temporal widths of the corresponding redshift bins, we include nominal detections from \cite{Agrawal+25} ($3\sigma$ upper limit), \cite{Hemmati17} ($1\sigma$), \cite{Yang+19} ($95\%$ confidence), and \cite{Yan20} ($1\sigma$) in pink, red, purple and black respectively.
        Grey fill regions display the $I_\cii$ history model space represented in current literature. We include three representative models: Dot-dashed gray line from \cite{padmanabhan+19}, Dashed gray line from \cite{yang+22}, dotted grey line from \cite{keenan20}. The dashed orange line shows the average \cii intensity/SFRD forecast in this work.}
    \label{fig:ICII}
\end{centering}
\end{figure*}

We apply Eq \ref{eq:CIIxGal} to determine TIM \cii x EDF-F cross-power spectrum models for each TIM redshift bin. Dividing the power (Eq \ref{eq:CIIxGal}) by the square root of the variance (Eq \ref{variance}) yields an SNR as a function of $k$ in bins of $d \ln(k) = 1$, as shown in the lower panels of Figure \ref{fig:constrainingPower}. In the upper panels, we present TIM's $3 \sigma$ detection threshold as well as our model for the TIM$\times$EDF-F intrinsic signal. 

We forecast that TIM will secure robust detections of the shot power-dominated regime ($k>1$ h Mpc$^{-1}$) in all four redshift bins in cross-correlation with the EDF-F galaxy distribution, with only slightly higher SNR in the two higher redshift spectral bins with respect to the shot regime. In TIM's two higher redshift bins, we predict successful detections of the transition from the \cii signal in the clustering regime ($k<1$ h Mpc$^{-1}$) to the shot regime. With decreasing $|k|$, the SNR suffers from two separate effects. First, the number of available mode samples decreases with $k$. Additionally, a higher proportion of available modes contain some directional component $k_{i} = 0$, and are therefore removed from our count. By $k\sim 0.1 \, h \, \mathrm{Mpc}^{-1}$, virtually no modes remain. If taken in aggregate (summed in quadrature), each of TIM's redshift bins should return a detection at a total SNR of at least 6.



The TIM SNR predictions presented in figure \ref{fig:constrainingPower} are from our semi-analytical halo formalism-based model, shown by the solid orange lines in figure \ref{fig:constrainingPower}. Our forecast agrees well with the most conservative representatives of the model space, which assume that \ICII closely traces SFR. Here again, we forecast that TIM$\times$\textit{Euclid} will generate definitive detections of the cross-shot power in the four highest $k$ bins extending across all redshift bins. In the two higher redshift bins, we expect to detect a roughly $1\sigma$ signal across all probable scales, extending to many $\sigma$ detections in the case of the more optimistic model space. 


In Figure \ref{fig:ICII} we report the inverse variance weighted average of \ICII. This is done individually for each of TIM's 4 redshift bins. 
Here we assume that the galaxy and \cii clustering bias factors follow the models in section
\ref{crossCorrChoices}. 
The upward arrows in Figure \ref{fig:ICII} present the $3 \sigma$ sensitivity threshold toward the \ICII signal. Here we predict that TIM will fully constrain the \ICII history across $0.5 \lesssim z \lesssim 1.7$. These measurements will provide an independent view of the total (obscured and unobscured) star formation rate density of the universe over a large fraction of cosmic time. In the grey-filled region we include the model space spanned by some commonly considered literature. (The discontinuity at $z\sim1$ is due to the fact that one model does not extend to $z=0$.) 

\subsection{Discussion}
\label{sec:Discussion}
\begin{figure}[t!]
\begin{centering}\includegraphics[width=0.48\textwidth]{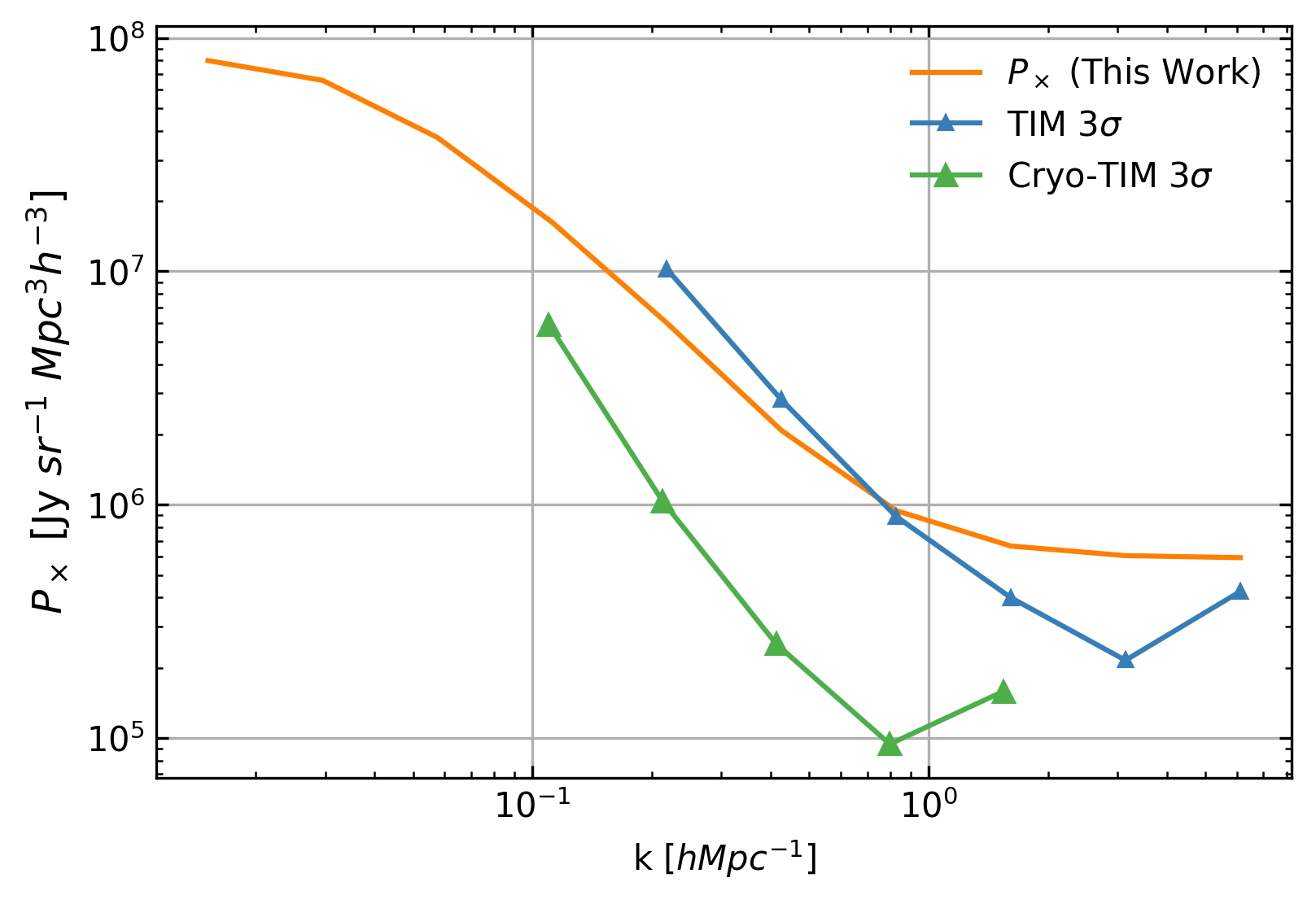}
        \caption{3$\sigma$ galaxy-cross-power-spectrum detection thresholds for TIM (blue) and a nominal next-generation TIM-like experiment, Cryo-TIM (green) in TIM's $1.33 < z < 1.66$ band. Assuming a 4 $\mathrm{deg}^2$ field and an atmosphere-limited, 4 Kelvin, 0.5m primary mirror for Cryo-TIM, and holding all other parameters constant between experiments.}
    \label{fig:TNG}
\end{centering}
\end{figure}

Achieving robust auto-power detections at higher redshift and building a deep and precise inventory of the SFRD history will require the implementation of surveys that are capable of both the depth and breadth of field to overcome cosmic variance uncertainties. This will require wider fields of view, which in turn require either increased detector hours or decreased atmospheric/instrumental loading of FIR detectors - ideally, both. In figure \ref{fig:TNG}, we compare the detection power of the TIM experiment as built, against a next-generation TIM-like experiment which simply replaces TIM's warm 2m mirror with a cold 0.5m mirror (see fig \ref{fig:NEIs} for the corresponding atmosphere-limited NEI curve), and assumes a 4 $\mathrm{deg}^2$ survey field; we hold all other experimental parameters constant, including the 200 hour survey time. By integrating a cold-mirror approach similar to the design of the \textit{PIPER} and \textit{EXCLAIM} \citep{piperOverview, exclaimOverview} balloon payloads, an $\sim$order of magnitude increase in sensitivity could be achieved.
A complementary paper, \citet{Agrawal+26}, further explores the prospect of galaxy cross-correlation with hypothetical successors to TIM (i.e., balloon and space far-infrared spectrometers). \citet{Agrawal+26} also forecasts the sensitivity of TIM in cross-correlation with other LIM tracers, which provide complementary information to the galaxy cross-correlation in both the clustering and shot regimes.

\section{Conclusion}
\label{sec:Conclusion}

\cii LIM-based constraints on the SFRD history are highly complementary to existing measurements. \cii can provide an unbiased accounting of dust-obscured star formation,  avoiding the dust corrections required of UV-based measurements. At the same time, intensity mapping constrains the faint end of the luminosity function, probing galaxies too faint to be directly detected in IR continuum surveys. Because the \LxG cross-shot power is sensitive to the coincident presence of star formation in both the LIM and galaxy surveys, the shot regime measurement will be particularly helpful in informing the level to which our galaxy catalogs undercount low mass and dust-obscured galaxies as a function of redshift.

Additionally, the wide-field nature of LIM -- especially with future mission concepts -- also gives rise to field sizes much larger than typical extra-galactic deep fields, providing a robust check on the field-to-field variance that is often a challenge for surveys with narrow field instruments like HST, JWST, and ALMA.

We have reviewed a simple formalism for computing the expected signal strength and error in a cross-power spectrum between a line intensity map and the overdensity field of a spectroscopic galaxy survey. We explored the expected SNR on this cross-correlation for the case of the imminent TIM \cii survey, taken with the upcoming Euclid Deep Field Fornax. We forecast that this approach will produce definitive constraints on the star formation history. TIM will confirm or rule out speculative models of the \ICII history, and will, at the very least, provide robust detections of the cross-power spectral shot signal during cosmic noon.


With the ability to push to deeper redshifts, near-future balloon experiments will be able to fill the $1<z<3$ gap where \cii must be studied from above the atmosphere. At greater redshifts, facilities like FYST equipped with instruments like EoR-Spec will be able to probe \cii out to the epoch of reionization. This will give us a wide-field, unbiased view of the cosmic SFR history, using a single, well-calibrated tracer, essentially covering from reionization to cosmic noon. 
This would enable cross-correlations of the \cii signal with nearly any other emission line across the vast majority of the universe's history, allowing astronomers to probe relative chemical abundances as well as the contributions of certain line species toward the aggregate emission of small, low-metallicity galaxies \citep{schaan&white21Applications}. 

Finally, TIM serves as a pathfinder mission with respect to both data analysis techniques and hardware development for upcoming missions. For instance, this study, and TIM's science goals, are deeply relevant and complementary to PRIMA (Probe far-Infrared Mission for Astrophysics), a NASA Astrophysics Probe Explorer concept for a 1.8 m, $\approx 4.5$ K cryogenic far-infrared ($\approx$ 24–261 \mum) observatory \citep{PRIMA_concept}. Similarly, as this first generation of LIM experiments come online, they enable not only a wealth of data, also a wealth of possibility for near-term growth in astrophysics and cosmology. 

\subsection*{Acknowledgments}
Work on the Terahertz Intensity Mapper is supported by the National Aeronautics and Space Administration under grants 80NSSC19K1242 and 80NSSC24K1881 issued through the Science Mission Directorate.

S.A. and J.A. note that this publication was made possible through the support of Grant 63040 from the John Templeton Foundation. The opinions expressed in this publication are those of the authors and do not necessarily reflect the views of the John Templeton Foundation. S.A.'s work was also supported by the Quad Fellowship.

Part of this research was carried out at the Jet Propulsion Laboratory, California Institute of Technology, under a contract with the National Aeronautics and Space Administration (80NM0018D0004).

\pagebreak

\appendix{}
\section{Surveys and Experiments}
\label{sec:GalSurvs}

\begin{table*}[t]
    \centering
    \begin{tabular}{c | c | c | c | c | c | c }
         \textbf{} & \textbf{$R_{\rm spec}$} ($\lambda / \Delta\lambda$) & \textbf{$L_{\rm lim}$} (Mag) & \textbf{$A_{\rm Surv}$} ($\mathrm{deg}^2$) & 
         $A_{\rm over}$ &
         $z_{\rm over}$ & 
         {Filter/Spectrometer} \\
         
         \hline
         \hline
         
         EDF-F & 450 & $\sim 24$ & 10 & 100\% & 100\% & Red \& Blue Grisms / H$\alpha$ \\
         \hline
         astroDEEP & 350-450 &  $\sim 24$ & 0.08 & 40\% & 100\% & Various NIR \\
         \hline
         Roman & 461 &  $\sim 24$ & 2000 & 100\% &  $\sim 50$\% &  G150 / H$\alpha$ \\
         \hline
         3D-DaSh & 350 & 24.74 & 1.35 & 0\% & 100\% & F160W\\
         
         \hline
         \hline
    \end{tabular}
    \caption{Parameters of four example galaxy surveys with which the TIM lineage of telescopes could cross-correlate. $L_{\rm lim}$ is the luminosity of a point source that would yield a $5\sigma$ detection. $A_{\rm over}$ and $z_{\rm over}$ reference the survey's field and redshift overlaps with TIM respectively. Filters and Spectrometers are referred to by the monikers set down in the literature (no attempt was made to standardize naming convention across experiments). }\vspace{-0.02in}
    \label{tab: galaxy_surveys}
\end{table*}


Cross-correlations with optical and infrared galaxy catalogues offer high SNR cross-checks on LIM. In the limit that $P_{\rm Gal}$ is dominated by its shot-noise term ($1/n_{\rm Gal}$), the SNR of a \LxG power spectrum is heavily dependent on the galaxy catalogue's number density. 
Therefore, the utility of a galaxy survey for LIM cross-correlations depends on the minimum detectable luminosity, $L_{\rm min}$, of
the galaxies in the traditional survey. A relatively low limiting luminosity is required to achieve a sufficiently high $n_{\rm gal}$ and thus mitigate shot-noise contributions to the cross-power spectrum variance. 
This is currently challenging at high redshifts, e.g. $z \gtrsim 2$, do to a lack of sufficiently deep galaxy surveys, but fairly routine at the redshifts of interest for TIM, $0.5 \lesssim z \lesssim 1.7$. 

Table \ref{tab: galaxy_surveys} serves as a quick reference and summary of the following discussion of some galaxy surveys which are potential cross-correlators for the TIM experiment (and future TIM-like experiments). Certain HST legacy catalogues offer sufficient $n_{\rm Gal}$ for cross-correlations, but suffer from sky coverage limitations.
The 3D-DASH catalogue \citep{mowla3dDASH} is a $\sim 1.43$ square degree survey of the equatorial COSMOS field, which entirely encompasses TIM's redshift range, and fully covers TIM's target field size. 
3D-DASH probed galaxies with luminosities down to $\sim 25th$ magnitude ($M_{AB} \approx 24.7$) for point sources. This catalogue \textit{would} be ideal for cross-correlations with the baseline, first-generation TIM experiment. As an Antarctic balloon experiment, TIM is unable to view COSMOS and other more northerly fields. A subsequent long-duration balloon flight at a higher latitude could make great use of both the 3D-DASH catalogue and the increase in the technological readiness level of necessary flight and detector hardware made by the TIM collaboration. 

Similarly, the ASTRODEEP-GS34 \citep{Merlin_2021} catalogue is an existing compilation of spectroscopic and photometric detections of galaxies in the GOOD-S field. ASTRODEEP / \textit{CANDELS} is one of the deepest and densest catalogues of spectroscopic galaxies currently available. However, the relatively small field of GOODS-S and the ASTRODEEP catalogue ($\approx 0.08$\degsq) imposes a steep limitation on its potential science yield in cross-correlation with LIM surveys, which typically aim to map regions of $>0.1$\degsq\ to avoid uncertainty from sample variance.  ASTRODEEP provides a promising proof of concept. A similar, but wider galaxy catalogue would yield even more valuable science data.

Ground-based, wide-field, spectrographic galaxy surveys have achieved impressive breadth in terms of survey area. Still, they generally lack the number density necessary for successful cross-correlations in $\sim \mathrm{deg}^2$ fields such as those targeted by TIM. 
Fortunately, upcoming and ongoing space telescopes will provide deep and wide-field spectroscopic galaxy catalogues.
For instance, NASA's Nancy Grace Roman Space Telescope, \textit{Roman}, is forecast to detect redshifted H$\alpha$ line emitting galaxies from $1 \lesssim z \lesssim 2$ in numbers comparable to the \textit{CANDELS} catalog across vastly larger regions of the sky. This will overlap in redshift with TIM's long wavelength module only \citep{wang22,pozetti16}. More relavent for TIM's purposes, given it's timeline and observing capabilities, is the recently launched \textit{Euclid} space telescope which will map the galaxy distribution at high sensitivity using the redshifted H$\alpha$ line \citep{euclid24}. 
\textit{Euclid}'s deep-field Fornax (EDF-F) is a $\sim 10$\degsq\ field positioned over Chandra deep-field south (100 times the field area over ASTRODEEP). This calibration field promises the most complete galaxy population for cross-correlation with the TIM \cii survey. While the \textit{Euclid} wide survey will employ only their red grism, EDF-F is being probed with both the experiment's red and blue grisms \citep{euclidNISP}.
Their combined wavelength sensitivity traces the H$\alpha$ line at $0.42 \lesssim z \lesssim 1.8$. \textit{Euclid} will probe this field down to an absolute magnitude of $ \gtrsim 24$ in both its wavelength bands. This will match the galaxy number density of ASTRODEEP at $z \sim 1$ and more than double the ASTRODEEP $n_{\rm Gal}$ by $z = 2$ \citep{pozetti16}. We will adopt EDF-F as our galaxy catalogue cross-correlator for the remainder of this work.






\clearpage

\bibliography{bib}

@article{Agrawal+25,
   title={Far-infrared lines hidden in archival deep multi-wavelength surveys: Limits on [CII]-158μm at z ~ 0.3 - 2.9},
   ISSN={1432-0746},
   url={http://dx.doi.org/10.1051/0004-6361/202556503},
   DOI={10.1051/0004-6361/202556503},
   journal={Astronomy \& Astrophysics},
   publisher={EDP Sciences},
   author={Agrawal, Shubh and Aguirre, James and Keenan, Ryan},
   year={2025},
   month=dec }

@article{Agrawal+26,
  title={Probing baryon statistics at cosmic noon with deep-field line intensity mapping: auto and cross spectra forecasts for \cii 158\mum, HI-21cm, CO, and H$\alpha$-galaxy counts at $z \sim 0.5 - 1.6$},
  author={Agrawal, Shubh and Aguirre, James and Keenan, Ryan and Bracks, Justin and Van Cuyck, Mathilde and Garcia, Karolina and Karoumpis, Christos and Zebrowski, Jessica and The TIM Collaboration},
  journal={},
  year={in prep.}
}

@ARTICLE{Amiri23,
       author = {{Amiri}, Mandana and {Bandura}, Kevin and {Chen}, Tianyue and {Deng}, Meiling and {Dobbs}, Matt and {Fandino}, Mateus and {Foreman}, Simon and {Halpern}, Mark and {Hill}, Alex S. and {Hinshaw}, Gary and {H{\"o}fer}, Carolin and {Kania}, Joseph and {Landecker}, T.~L. and {MacEachern}, Joshua and {Masui}, Kiyoshi and {Mena-Parra}, Juan and {Milutinovic}, Nikola and {Mirhosseini}, Arash and {Newburgh}, Laura and {Ordog}, Anna and {Pen}, Ue-Li and {Pinsonneault-Marotte}, Tristan and {Polzin}, Ava and {Reda}, Alex and {Renard}, Andre and {Shaw}, J. Richard and {Siegel}, Seth R. and {Singh}, Saurabh and {Vanderlinde}, Keith and {Wang}, Haochen and {Wiebe}, Donald V. and {Wulf}, Dallas and {CHIME Collaboration}},
        title = "{Detection of Cosmological 21 cm Emission with the Canadian Hydrogen Intensity Mapping Experiment}",
      journal = {\apj},
         year = 2023,
        month = apr,
       volume = {947},
       number = {1},
          eid = {16},
        pages = {16},
          doi = {10.3847/1538-4357/acb13f},
archivePrefix = {arXiv},
       eprint = {2202.01242},
 primaryClass = {astro-ph.CO},
       adsurl = {https://ui.adsabs.harvard.edu/abs/2023ApJ...947...16A}
}

@misc{atmospheric_model,
  author       = {Paine, Scott},
  title        = {The am atmospheric model},
  month        = sep,
  year         = 2024,
  publisher    = {Zenodo},
  version      = {14.0},
  doi          = {10.5281/zenodo.13748391},
  url          = {https://doi.org/10.5281/zenodo.13748391},
}

@article{bernal&Kovetz22,
  title={Line-intensity mapping: theory review with a focus on star-formation lines},
  author={Bernal, Jos{\'e} Luis and Kovetz, Ely D},
  journal={The Astronomy and Astrophysics Review},
  volume={30},
  number={1},
  pages={5},
  year={2022},
  publisher={Springer}
}

@article{cunnington2023h,
  title={H i intensity mapping with MeerKAT: power spectrum detection in cross-correlation with WiggleZ galaxies},
  author={Cunnington, Steven and Li, Yichao and Santos, Mario G and Wang, Jingying and Carucci, Isabella P and Irfan, Melis O and Pourtsidou, Alkistis and Spinelli, Marta and Wolz, Laura and Soares, Paula S and others},
  journal={Monthly Notices of the Royal Astronomical Society},
  volume={518},
  number={4},
  pages={6262--6272},
  year={2023},
  publisher={Oxford University Press}
}

@article{pen2009,
  title={First detection of cosmic structure in the 21-cm intensity field},
  author={Pen, Ue-Li and Staveley-Smith, Lister and Peterson, Jeffrey B and Chang, Tzu-Ching},
  journal={Monthly Notices of the Royal Astronomical Society: Letters},
  volume={394},
  number={1},
  pages={L6--L10},
  year={2009},
  publisher={Blackwell Science Ltd Oxford, UK}
}

@article{anderson2018,
  title={Low-amplitude clustering in low-redshift 21-cm intensity maps cross-correlated with 2dF galaxy densities},
  author={Anderson, CJ and Luciw, NJ and Li, Y-C and Kuo, CY and Yadav, J and Masui, KW and Chang, TC and Chen, X and Oppermann, N and Liao, YW and others},
  journal={Monthly Notices of the Royal Astronomical Society},
  volume={476},
  number={3},
  pages={3382--3392},
  year={2018},
  publisher={Oxford University Press}
}

@article{wolz2022,
  title={H i constraints from the cross-correlation of eBOSS galaxies and Green Bank Telescope intensity maps},
  author={Wolz, Laura and Pourtsidou, Alkistis and Masui, Kiyoshi W and Chang, Tzu-Ching and Bautista, Julian E and M{\"u}ller, Eva-Maria and Avila, Santiago and Bacon, David and Percival, Will J and Cunnington, Steven and others},
  journal={Monthly Notices of the Royal Astronomical Society},
  volume={510},
  number={3},
  pages={3495--3511},
  year={2022},
  publisher={Oxford University Press}
}

@ARTICLE{behroozi19,
       author = {{Behroozi}, Peter and {Wechsler}, Risa H. and {Hearin}, Andrew P. and {Conroy}, Charlie},
        title = "{UNIVERSEMACHINE: The correlation between galaxy growth and dark matter halo assembly from z = 0-10}",
      journal = {\mnras},
         year = 2019,
        month = sep,
       volume = {488},
       number = {3},
        pages = {3143-3194},
          doi = {10.1093/mnras/stz1182},
archivePrefix = {arXiv},
       eprint = {1806.07893},
 primaryClass = {astro-ph.GA},
       adsurl = {https://ui.adsabs.harvard.edu/abs/2019MNRAS.488.3143B}
}

@ARTICLE{bonato19,
       author = {{Bonato}, Matteo and {De Zotti}, Gianfranco and {Leisawitz}, David and {Negrello}, Mattia and {Massardi}, Marcella and {Baronchelli}, Ivano and {Cai}, Zhen-Yi and {Bradford}, Charles M. and {Pope}, Alexandra and {Murphy}, Eric J. and {Armus}, Lee and {Cooray}, Asantha},
        title = "{Origins Space Telescope: Predictions for far-IR spectroscopic surveys}",
      journal = {\pasa},
         year = 2019,
        month = apr,
       volume = {36},
          eid = {e017},
        pages = {e017},
          doi = {10.1017/pasa.2019.8},
archivePrefix = {arXiv},
       eprint = {1903.00946},
 primaryClass = {astro-ph.GA},
       adsurl = {https://ui.adsabs.harvard.edu/abs/2019PASA...36...17B}
}

@article{casey14,
	Adsurl = {http://adsabs.harvard.edu/abs/2014PhR...541...45C},
	Archiveprefix = {arXiv},
	Author = {{Casey}, C.~M. and {Narayanan}, D. and {Cooray}, A.},
	Doi = {10.1016/j.physrep.2014.02.009},
	Eprint = {1402.1456},
	Journal = {\physrep},
	Month = aug,
	Pages = {45-161},
	Title = {{Dusty star-forming galaxies at high redshift}},
	Volume = {541},
	Year = {2014}}

@article{chang2010,
  title={An intensity map of hydrogen 21-cm emission at redshift z≈ 0.8},
  author={Chang, Tzu-Ching and Pen, Ue-Li and Bandura, Kevin and Peterson, Jeffrey B},
  journal={Nature},
  volume={466},
  number={7305},
  pages={463--465},
  year={2010},
  publisher={Nature Publishing Group UK London}
}

@misc{changLidz26,
      title={Line-Intensity Mapping}, 
      author={Tzu-Ching Chang and Adam Lidz},
      year={2026},
      eprint={2602.03011},
      archivePrefix={arXiv},
      primaryClass={astro-ph.CO},
      url={https://arxiv.org/abs/2602.03011}, 
}

@article{Choi23CCAT,
	author = {Choi, Steve and Collaboration, CCAT-prime},
	journal = {Bulletin of the AAS},
	number = {2},
	year = {2023},
	month = {jan 31},
	note = {https://baas.aas.org/pub/2023n2i346p02},
	publisher = {},
	title = {CCAT-prime: Science {Goals} and {Forecasts} with {Prime}-{Cam} on the {Fred} {Young} {Submillimeter} {Telescope}},
	volume = {55},
}

@inproceedings{Crites22,
author = {Abigail T. Crites},
title = {{TIME, the Tomographic Ionized Carbon Intensity Mapping Experiment: an update on design, characterization, and data from the 2022 commissioning observations}},
volume = {PC12190},
booktitle = {Millimeter, Submillimeter, and Far-Infrared Detectors and Instrumentation for Astronomy XI},
editor = {Jonas Zmuidzinas and Jian-Rong Gao},
organization = {International Society for Optics and Photonics},
publisher = {SPIE},
pages = {PC121900I},
year = {2022},
doi = {10.1117/12.2630647},
URL = {https://doi.org/10.1117/12.2630647}
}

@ARTICLE{Cunnington23,
       author = {{Cunnington}, Steven and {Li}, Yichao and {Santos}, Mario G. and {Wang}, Jingying and {Carucci}, Isabella P. and {Irfan}, Melis O. and {Pourtsidou}, Alkistis and {Spinelli}, Marta and {Wolz}, Laura and {Soares}, Paula S. and {Blake}, Chris and {Bull}, Philip and {Engelbrecht}, Brandon and {Fonseca}, Jos{\'e} and {Grainge}, Keith and {Ma}, Yin-Zhe},
        title = "{H I intensity mapping with MeerKAT: power spectrum detection in cross-correlation with WiggleZ galaxies}",
      journal = {\mnras},
         year = 2023,
        month = feb,
       volume = {518},
       number = {4},
        pages = {6262-6272},
          doi = {10.1093/mnras/stac3060},
archivePrefix = {arXiv},
       eprint = {2206.01579},
 primaryClass = {astro-ph.CO},
       adsurl = {https://ui.adsabs.harvard.edu/abs/2023MNRAS.518.6262C}
}

@BOOK{draine11,
       author = {{Draine}, Bruce T.},
        title = "{Physics of the Interstellar and Intergalactic Medium}",
         year = 2011,
       adsurl = {https://ui.adsabs.harvard.edu/abs/2011piim.book.....D}
}

@ARTICLE{Durkalec15,
       author = {{Durkalec}, A. and {Le F{\`e}vre}, O. and {Pollo}, A. and {de la Torre}, S. and {Cassata}, P. and {Garilli}, B. and {Le Brun}, V. and {Lemaux}, B.~C. and {Maccagni}, D. and {Pentericci}, L. and {Tasca}, L.~A.~M. and {Thomas}, R. and {Vanzella}, E. and {Zamorani}, G. and {Zucca}, E. and {Amor{\'\i}n}, R. and {Bardelli}, S. and {Cassar{\`a}}, L.~P. and {Castellano}, M. and {Cimatti}, A. and {Cucciati}, O. and {Fontana}, A. and {Giavalisco}, M. and {Grazian}, A. and {Hathi}, N.~P. and {Ilbert}, O. and {Paltani}, S. and {Ribeiro}, B. and {Schaerer}, D. and {Scodeggio}, M. and {Sommariva}, V. and {Talia}, M. and {Tresse}, L. and {Vergani}, D. and {Capak}, P. and {Charlot}, S. and {Contini}, T. and {Cuby}, J.~G. and {Dunlop}, J. and {Fotopoulou}, S. and {Koekemoer}, A. and {L{\'o}pez-Sanjuan}, C. and {Mellier}, Y. and {Pforr}, J. and {Salvato}, M. and {Scoville}, N. and {Taniguchi}, Y. and {Wang}, P.~W.},
        title = "{Evolution of clustering length, large-scale bias, and host halo mass at 2 < z < 5 in the VIMOS Ultra Deep Survey (VUDS){\ensuremath{\star}}}",
      journal = {\aap},
         year = 2015,
        month = nov,
       volume = {583},
          eid = {A128},
        pages = {A128},
          doi = {10.1051/0004-6361/201425343},
archivePrefix = {arXiv},
       eprint = {1411.5688},
 primaryClass = {astro-ph.CO},
       adsurl = {https://ui.adsabs.harvard.edu/abs/2015A&A...583A.128D}
}

@misc{euclidQ1Spec,
      title={Euclid Quick Data Release (Q1) -- Characteristics and limitations of the spectroscopic measurements}, 
      author={V. {Le Brun} and M. Bethermin and M. Moresco and D. Vibert and D. Vergani and C. Surace and G. Zamorani and A. Allaoui and T. Bedrine and P. -Y. Chabaud and G. Daste and F. Dufresne and M. Gray and E. Rossetti and Y. Copin and S. Conseil and E. Maiorano and Z. Mao and E. Palazzi and L. Pozzetti and S. Quai and C. Scarlata and M. Talia and H. M. Courtois and L. Guzzo and B. Kubik and A. M. C. Le Brun and J. A. Peacock and D. Scott and D. Bagot and A. Basset and P. Casenove and R. Gimenez and G. Libet and M. Ruffenach and N. Aghanim and B. Altieri and A. Amara and S. Andreon and N. Auricchio and H. Aussel and C. Baccigalupi and M. Baldi and A. Balestra and S. Bardelli and P. Battaglia and A. Biviano and A. Bonchi and D. Bonino and E. Branchini and M. Brescia and J. Brinchmann and A. Caillat and S. Camera and G. Cañas-Herrera and V. Capobianco and C. Carbone and J. Carretero and S. Casas and F. J. Castander and G. Castignani and S. Cavuoti and K. C. Chambers and A. Cimatti and C. Colodro-Conde and G. Congedo and C. J. Conselice and L. Conversi and A. Costille and F. Courbin and J. -G. Cuby and A. Da Silva and H. Degaudenzi and S. de la Torre and G. De Lucia and A. M. Di Giorgio and H. Dole and M. Douspis and F. Dubath and X. Dupac and S. Dusini and A. Ealet and S. Escoffier and M. Fabricius and M. Farina and R. Farinelli and F. Faustini and S. Ferriol and S. Fotopoulou and N. Fourmanoit and M. Frailis and E. Franceschi and M. Fumana and S. Galeotta and K. George and W. Gillard and B. Gillis and C. Giocoli and J. Gracia-Carpio and B. R. Granett and A. Grazian and F. Grupp and S. V. H. Haugan and J. Hoar and H. Hoekstra and W. Holmes and F. Hormuth and A. Hornstrup and P. Hudelot and K. Jahnke and M. Jhabvala and B. Joachimi and E. Keihänen and S. Kermiche and A. Kiessling and M. Kümmel and M. Kunz and H. Kurki-Suonio and Q. Le Boulc'h and D. Le Mignant and S. Ligori and P. B. Lilje and V. Lindholm and I. Lloro and G. Mainetti and D. Maino and O. Mansutti and S. Marcin and O. Marggraf and M. Martinelli and N. Martinet and F. Marulli and R. Massey and S. Maurogordato and E. Medinaceli and S. Mei and M. Melchior and Y. Mellier and M. Meneghetti and E. Merlin and G. Meylan and A. Mora and L. Moscardini and R. Nakajima and C. Neissner and R. C. Nichol and S. -M. Niemi and J. W. Nightingale and C. Padilla and S. Paltani and F. Pasian and K. Pedersen and W. J. Percival and V. Pettorino and S. Pires and G. Polenta and M. Poncet and L. A. Popa and F. Raison and R. Rebolo and A. Renzi and J. Rhodes and G. Riccio and E. Romelli and M. Roncarelli and R. Saglia and Z. Sakr and D. Sapone and B. Sartoris and M. Sauvage and J. A. Schewtschenko and M. Schirmer and P. Schneider and T. Schrabback and M. Scodeggio and A. Secroun and G. Seidel and M. Seiffert and C. Sirignano and G. Sirri and L. Stanco and J. Steinwagner and P. Tallada-Crespí and A. N. Taylor and H. I. Teplitz and I. Tereno and N. Tessore and S. Toft and R. Toledo-Moreo and F. Torradeflot and I. Tutusaus and L. Valenziano and J. Valiviita and T. Vassallo and G. Verdoes Kleijn and A. Veropalumbo and Y. Wang and J. Weller and A. Zacchei and F. M. Zerbi and I. A. Zinchenko and E. Zucca and V. Allevato and M. Ballardini and M. Bolzonella and E. Bozzo and C. Burigana and R. Cabanac and A. Cappi and D. Di Ferdinando and J. A. Escartin Vigo and G. Fabbian and L. Gabarra and W. G. Hartley and J. Martín-Fleitas and S. Matthew and M. Maturi and N. Mauri and R. B. Metcalf and A. Pezzotta and M. Pöntinen and C. Porciani and I. Risso and V. Scottez and M. Sereno and M. Tenti and M. Viel and M. Wiesmann and Y. Akrami and S. Alvi and I. T. Andika and S. Anselmi and M. Archidiacono and F. Atrio-Barandela and S. Avila and M. Bella and P. Bergamini and D. Bertacca and L. Blot and S. Borgani and M. L. Brown and S. Bruton and A. Calabro and B. Camacho Quevedo and F. Caro and C. S. Carvalho and T. Castro and Y. Charles and R. Chary and F. Cogato and A. R. Cooray and O. Cucciati and S. Davini and F. De Paolis and G. Desprez and A. Díaz-Sánchez and J. J. Diaz and S. Di Domizio and J. M. Diego and P. Dimauro and P. -A. Duc and A. Enia and Y. Fang and A. M. N. Ferguson and A. G. Ferrari and A. Finoguenov and A. Fontana and A. Franco and K. Ganga and J. García-Bellido and T. Gasparetto and V. Gautard and E. Gaztanaga and F. Giacomini and F. Gianotti and G. Gozaliasl and A. Gregorio and M. Guidi and C. M. Gutierrez and A. Hall and C. Hernández-Monteagudo and H. Hildebrandt and J. Hjorth and J. J. E. Kajava and Y. Kang and V. Kansal and D. Karagiannis and K. Kiiveri and C. C. Kirkpatrick and S. Kruk and L. Legrand and M. Lembo and F. Lepori and G. F. Lesci and J. Lesgourgues and L. Leuzzi and T. I. Liaudat and S. J. Liu and A. Loureiro and J. Macias-Perez and M. Magliocchetti and E. A. Magnier and C. Mancini and F. Mannucci and R. Maoli and C. J. A. P. Martins and L. Maurin and M. Miluzio and P. Monaco and A. Montoro and C. Moretti and G. Morgante and S. Nadathur and K. Naidoo and A. Navarro-Alsina and S. Nesseris and F. Passalacqua and K. Paterson and L. Patrizii and A. Pisani and D. Potter and M. Radovich and P. -F. Rocci and S. Sacquegna and M. Sahlén and D. B. Sanders and E. Sarpa and A. Schneider and D. Sciotti and E. Sellentin and F. Shankar and L. C. Smith and K. Tanidis and G. Testera and R. Teyssier and S. Tosi and A. Troja and M. Tucci and C. Valieri and A. Venhola and G. Verza and P. Vielzeuf and N. A. Walton and J. R. Weaver and L. Zalesky and J. G. Sorce},
      year={2025},
      eprint={2503.15308},
      archivePrefix={arXiv},
      primaryClass={astro-ph.CO},
      url={https://arxiv.org/abs/2503.15308}, 
}

@misc{euclidNISP,
      title={Euclid. III. The NISP Instrument}, 
      author={K. Jahnke and W. Gillard and M. Schirmer and A. Ealet and T. Maciaszek and E. Prieto and R. Barbier and C. Bonoli and L. Corcione and S. Dusini and F. Grupp and F. Hormuth and S. Ligori and L. Martin and G. Morgante and C. Padilla and R. Toledo-Moreo and M. Trifoglio and L. Valenziano and R. Bender and F. J. Castander and B. Garilli and P. B. Lilje and H. -W. Rix and N. Auricchio and A. Balestra and J. -C. Barriere and P. Battaglia and M. Berthe and C. Bodendorf and T. Boenke and W. Bon and A. Bonnefoi and A. Caillat and V. Capobianco and M. Carle and R. Casas and H. Cho and A. Costille and F. Ducret and S. Ferriol and E. Franceschi and J. -L. Gimenez and W. Holmes and A. Hornstrup and M. Jhabvala and R. Kohley and B. Kubik and R. Laureijs and D. Le Mignant and I. Lloro and E. Medinaceli and Y. Mellier and G. Polenta and G. D. Racca and A. Renzi and J. -C. Salvignol and A. Secroun and G. Seidel and M. Seiffert and C. Sirignano and G. Sirri and P. Strada and G. Smadja and L. Stanco and S. Wachter and S. Anselmi and E. Borsato and L. Caillat and F. Cogato and C. Colodro-Conde and P. -E. Crouzet and V. Conforti and M. D'Alessandro and Y. Copin and J. -C. Cuillandre and J. E. Davies and S. Davini and A. Derosa and J. J. Diaz and S. Di Domizio and D. Di Ferdinando and R. Farinelli and A. G. Ferrari and F. Fornari and L. Gabarra and C. M. Gutierrez and F. Giacomini and P. Lagier and F. Gianotti and O. Krause and F. Madrid and F. Laudisio and J. Macias-Perez and G. Naletto and M. Niclas and J. Marpaud and N. Mauri and R. da Silva and F. Passalacqua and K. Paterson and L. Patrizii and I. Risso and B. G. B. Solheim and M. Scodeggio and P. Stassi and J. Steinwagner and M. Tenti and G. Testera and R. Travaglini and S. Tosi and A. Troja and O. Tubio and C. Valieri and C. Vescovi and S. Ventura and N. Aghanim and B. Altieri and A. Amara and J. Amiaux and S. Andreon and H. Aussel and M. Baldi and S. Bardelli and A. Basset and A. Bonchi and D. Bonino and E. Branchini and M. Brescia and J. Brinchmann and S. Camera and C. Carbone and V. F. Cardone and J. Carretero and S. Casas and M. Castellano and S. Cavuoti and P. -Y. Chabaud and A. Cimatti and G. Congedo and C. J. Conselice and L. Conversi and F. Courbin and H. M. Courtois and M. Cropper and J. -G. Cuby and A. Da Silva and H. Degaudenzi and A. M. Di Giorgio and J. Dinis and M. Douspis and F. Dubath and C. A. J. Duncan and X. Dupac and M. Fabricius and M. Farina and S. Farrens and F. Faustini and P. Fosalba and S. Fotopoulou and N. Fourmanoit and M. Frailis and P. Franzetti and S. Galeotta and B. Gillis and C. Giocoli and P. Gómez-Alvarez and B. R. Granett and A. Grazian and L. Guzzo and M. Hailey and S. V. H. Haugan and J. Hoar and H. Hoekstra and I. Hook and P. Hudelot and B. Joachimi and E. Keihänen and S. Kermiche and A. Kiessling and M. Kilbinger and T. Kitching and M. Kümmel and M. Kunz and H. Kurki-Suonio and O. Lahav and V. Lindholm and J. Lorenzo Alvarez and D. Maino and E. Maiorano and O. Mansutti and O. Marggraf and K. Markovic and J. Martignac and N. Martinet and F. Marulli and R. Massey and D. C. Masters and S. Maurogordato and H. J. McCracken and S. Mei and M. Melchior and M. Meneghetti and E. Merlin and G. Meylan and J. J. Mohr and M. Moresco and L. Moscardini and R. Nakajima and R. C. Nichol and S. -M. Niemi and T. Nutma and K. Paech and S. Paltani and F. Pasian and J. A. Peacock and K. Pedersen and W. J. Percival and V. Pettorino and S. Pires and M. Poncet and L. A. Popa and L. Pozzetti and F. Raison and R. Rebolo and A. Refregier and J. Rhodes and G. Riccio and E. Romelli and M. Roncarelli and C. Rosset and E. Rossetti and H. J. A. Rottgering and R. Saglia and D. Sapone and M. Sauvage and R. Scaramella and P. Schneider and T. Schrabback and S. Serrano and P. Tallada-Crespí and D. Tavagnacco and A. N. Taylor and H. I. Teplitz and I. Tereno and F. Torradeflot and I. Tutusaus and T. Vassallo and G. Verdoes Kleijn and A. Veropalumbo and D. Vibert and Y. Wang and J. Weller and A. Zacchei and G. Zamorani and F. M. Zerbi and J. Zoubian and E. Zucca and P. N. Appleton and C. Baccigalupi and A. Biviano and M. Bolzonella and A. Boucaud and E. Bozzo and C. Burigana and M. Calabrese and P. Casenove and M. Crocce and G. De Lucia and J. A. Escartin Vigo and G. Fabbian and F. Finelli and K. George and J. Gracia-Carpio and S. Ilić and P. Liebing and C. Liu and G. Mainetti and S. Marcin and M. Martinelli and P. W. Morris and C. Neissner and A. Pezzotta and M. Pöntinen and C. Porciani and Z. Sakr and V. Scottez and E. Sefusatti and M. Viel and M. Wiesmann and Y. Akrami and V. Allevato and E. Aubourg and M. Ballardini and D. Bertacca and M. Bethermin and A. Blanchard and L. Blot and S. Borgani and A. S. Borlaff and S. Bruton and R. Cabanac and A. Calabro and G. Calderone and G. Canas-Herrera and A. Cappi and C. S. Carvalho and G. Castignani and T. Castro and K. C. Chambers and Y. Charles and R. Chary and J. Colbert and S. Contarini and T. Contini and A. R. Cooray and M. Costanzi and O. Cucciati and B. De Caro and S. de la Torre and G. Desprez and A. Díaz-Sánchez and H. Dole and S. Escoffier and P. G. Ferreira and I. Ferrero and A. Finoguenov and A. Fontana and K. Ganga and J. García-Bellido and V. Gautard and E. Gaztanaga and G. Gozaliasl and A. Gregorio and A. Hall and W. G. Hartley and S. Hemmati and H. Hildebrandt and J. Hjorth and S. Hosseini and M. Huertas-Company and O. Ilbert and J. Jacobson and S. Joudaki and J. J. E. Kajava and V. Kansal and D. Karagiannis and C. C. Kirkpatrick and F. Lacasa and V. Le Brun and J. Le Graet and L. Legrand and G. Libet and S. J. Liu and A. Loureiro and M. Magliocchetti and C. Mancini and F. Mannucci and R. Maoli and C. J. A. P. Martins and S. Matthew and L. Maurin and C. J. R. McPartland and R. B. Metcalf and M. Migliaccio and M. Miluzio and P. Monaco and C. Moretti and S. Nadathur and L. Nicastro and Nicholas A. Walton and J. Odier and M. Oguri and V. Popa and D. Potter and A. Pourtsidou and P. -F. Rocci and R. P. Rollins and B. Rusholme and M. Sahlén and A. G. Sánchez and C. Scarlata and J. Schaye and J. A. Schewtschenko and A. Schneider and M. Schultheis and M. Sereno and F. Shankar and A. Shulevski and G. Sikkema and A. Silvestri and P. Simon and A. Spurio Mancini and J. Stadel and S. A. Stanford and K. Tanidis and C. Tao and N. Tessore and R. Teyssier and S. Toft and M. Tucci and J. Valiviita and D. Vergani and F. Vernizzi and G. Verza and P. Vielzeuf and J. R. Weaver and L. Zalesky and I. A. Zinchenko and M. Archidiacono and F. Atrio-Barandela and C. L. Bennett and T. Bouvard and F. Caro and S. Conseil and P. Dimauro and P. -A. Duc and Y. Fang and A. M. N. Ferguson and T. Gasparetto and I. Kova{č}ić and S. Kruk and A. M. C. Le Brun and T. I. Liaudat and A. Montoro and A. Mora and C. Murray and L. Pagano and D. Paoletti and M. Radovich and E. Sarpa and E. Tommasi and A. Viitanen and J. Lesgourgues and M. E. Levi and J. Martín-Fleitas},
      year={2024},
      eprint={2405.13493},
      archivePrefix={arXiv},
      primaryClass={astro-ph.IM},
      url={https://arxiv.org/abs/2405.13493}, 
}

@inproceedings{Essinger_Hileman20Exclaim,
   title={Optical Design of the Experiment for Cryogenic Large-Aperture Intensity Mapping (EXCLAIM)},
   url={http://dx.doi.org/10.1117/12.2576254},
   DOI={10.1117/12.2576254},
   booktitle={Millimeter, Submillimeter, and Far-Infrared Detectors and Instrumentation for Astronomy X},
   publisher={SPIE},
   author={Essinger-Hileman, Thomas M. and Oxholm, Trevor M. and Siebert, Gage L. and Ade, Peter A. and Anderson, Christopher J. and Barlis, Alyssa and Barrentine, Emily M. and Beeman, Jeffrey and Bellis, Nicholas G. and Breysse, Patrick C. and Bolatto, Alberto D. and Bulcha, Berhanu T. and Cataldo, Giuseppe and Connors, Jake A. and Cursey, Paul W. and Ehsan, Negar and Fernandez, Lee-Roger and Glenn, Jason and Golec, Joseph E. and Hays-Wehle, James and Hess, Larry and Jahromi, Amir E. and Kimball, Mark O. and Kogut, Alan and Lowe, Luke N. and Mauskopf, Phil and McMahon, Jeffrey and Mirzaei, Mona and Moseley, Harvey and Mugge-Durum, Jonas and Noroozian, Omid and Pen, Ue-Li and Pullen, Anthony R. and Rodriguez, Samelys and Shire, Konrad and Sinclair, Adrian and Somerville, Rachel S. and Stevenson, Thomas R. and Switzer, Eric R. and Timbie, Peter and Tucker, Carole and Visbal, Eli and Volpert, Carolyn G. and Wollack, Edward J. and Yang, Shengqi},
   editor={Zmuidzinas, Jonas and Gao, Jian-Rong},
   year={2020},
   month=dec }

@inproceedings{exclaimOverview,
  title={Overview and status of EXCLAIM, the experiment for cryogenic large-aperture intensity mapping},
  author={Cataldo, Giuseppe and Ade, Peter AR and Anderson, Christopher J and Barlis, Alyssa and Barrentine, Emily M and Bellis, Nicholas G and Bolatto, Alberto D and Breysse, Patrick C and Bulcha, Berhanu T and Connors, Jake A and others},
  booktitle={Ground-based and Airborne Telescopes VIII},
  volume={11445},
  pages={469--479},
  year={2020},
  organization={SPIE}
}

@ARTICLE{hauser01,
   author = {{Hauser}, M.~G. and {Dwek}, E.},
    title = "{The Cosmic Infrared Background: Measurements and Implications}",
  journal = {\araa},
   eprint = {arXiv:astro-ph/0105539},
     year = 2001,
   volume = 39,
    pages = {249-307},
      doi = {10.1146/annurev.astro.39.1.249},
   adsurl = {http://adsabs.harvard.edu/abs/2001ARA\%26A..39..249H}
}

@article{lagache05,
	Adsurl = {http://adsabs.harvard.edu/abs/2005ARA%26A..43..727L},
	Author = {{Lagache}, G. and {Puget}, J.-L. and {Dole}, H.},
	Doi = {10.1146/annurev.astro.43.072103.150606},
	Eprint = {arXiv:astro-ph/0507298},
	Journal = {\araa},
	Month = sep,
	Pages = {727-768},
	Title = {{Dusty Infrared Galaxies: Sources of the Cosmic Infrared Background}},
	Volume = {43},
	Year = {2005}}

@ARTICLE{hollenbach97,
       author = {{Hollenbach}, D.~J. and {Tielens}, A.~G.~G.~M.},
        title = "{Dense Photodissociation Regions (PDRs)}",
      journal = {\araa},
         year = 1997,
        month = jan,
       volume = {35},
        pages = {179-216},
          doi = {10.1146/annurev.astro.35.1.179},
       adsurl = {https://ui.adsabs.harvard.edu/abs/1997ARA&A..35..179H}
}

@article{hollenbach99,
	author = {{Hollenbach}, D.~J. and {Tielens}, A.~G.~G.~M.},
	Doi = {10.1103/RevModPhys.71.173},
	Journal = {Reviews of Modern Physics},
	Month = jan,
	Pages = {173-230},
	Title = "{Photodissociation regions in the interstellar medium of galaxies}",
	Volume = {71},
	Year = {1999}
}

@ARTICLE{Meixner19,
       author = {{Meixner}, M. and {Cooray}, A. and {Leisawitz}, D. and {Staguhn}, J. and {Armus}, L. and {Battersby}, C. and {Bauer}, J. and {Bergin}, E. and {Bradford}, C.~M. and {Ennico-Smith}, K. and {Fortney}, J. and {Kataria}, T. and {Melnick}, G. and {Milam}, S. and {Narayanan}, D. and {Padgett}, D. and {Pontoppidan}, K. and {Pope}, A. and {Roellig}, T. and {Sandstrom}, K. and {Stevenson}, K. and {Su}, K. and {Vieira}, J. and {Wright}, E. and {Zmuidzinas}, J. and {Sheth}, K. and {Benford}, D. and {Mamajek}, E.~E. and {Neff}, S. and {De Beck}, E. and {Gerin}, M. and {Helmich}, F. and {Sakon}, I. and {Scott}, D. and {Vavrek}, R. and {Wiedner}, M. and {Carey}, S. and {Burgarella}, D. and {Moseley}, S.~H. and {Amatucci}, E. and {Carter}, R.~C. and {DiPirro}, M. and {Wu}, C. and {Beaman}, B. and {Beltran}, P. and {Bolognese}, J. and {Bradley}, D. and {Corsetti}, J. and {D'Asto}, T. and {Denis}, K. and {Derkacz}, C. and {Earle}, C.~P. and {Fantano}, L.~G. and {Folta}, D. and {Gavares}, B. and {Generie}, J. and {Hilliard}, L. and {Howard}, J.~M. and {Jamil}, A. and {Jamison}, T. and {Lynch}, C. and {Martins}, G. and {Petro}, S. and {Ramspacher}, D. and {Rao}, A. and {Sandin}, C. and {Stoneking}, E. and {Tompkins}, S. and {Webster}, C.},
        title = "{Origins Space Telescope Mission Concept Study Report}",
      journal = {arXiv e-prints},
         year = 2019,
        month = dec,
          eid = {arXiv:1912.06213},
        pages = {arXiv:1912.06213},
          doi = {10.48550/arXiv.1912.06213},
archivePrefix = {arXiv},
       eprint = {1912.06213},
 primaryClass = {astro-ph.IM},
       adsurl = {https://ui.adsabs.harvard.edu/abs/2019arXiv191206213M}
}

@ARTICLE{fonseca17,
       author = {{Fonseca}, Jos{\'e} and {Silva}, Marta B. and {Santos}, M{\'a}rio G. and {Cooray}, Asantha},
        title = "{Cosmology with intensity mapping techniques using atomic and molecular lines}",
      journal = {\mnras},
         year = 2017,
        month = jan,
       volume = {464},
       number = {2},
        pages = {1948-1965},
          doi = {10.1093/mnras/stw2470},
archivePrefix = {arXiv},
       eprint = {1607.05288},
 primaryClass = {astro-ph.CO},
       adsurl = {https://ui.adsabs.harvard.edu/abs/2017MNRAS.464.1948F}
}

@ARTICLE{moradinezhaddizgah19,
       author = {{Moradinezhad Dizgah}, Azadeh and {Keating}, Garrett K.},
        title = "{Line Intensity Mapping with [C II] and CO(1-0) as Probes of Primordial Non-Gaussianity}",
      journal = {\apj},
         year = 2019,
        month = feb,
       volume = {872},
       number = {2},
          eid = {126},
        pages = {126},
          doi = {10.3847/1538-4357/aafd36},
archivePrefix = {arXiv},
       eprint = {1810.02850},
 primaryClass = {astro-ph.CO},
       adsurl = {https://ui.adsabs.harvard.edu/abs/2019ApJ...872..126M}
}

@ARTICLE{karkare22,
       author = {{Karkare}, Kirit S. and {Moradinezhad Dizgah}, Azadeh and {Keating}, Garrett K. and {Breysse}, Patrick and {Chung}, Dongwoo T.},
        title = "{Snowmass 2021 Cosmic Frontier White Paper: Cosmology with Millimeter-Wave Line Intensity Mapping}",
      journal = {arXiv e-prints},
         year = 2022,
        month = mar,
          eid = {arXiv:2203.07258},
        pages = {arXiv:2203.07258},
          doi = {10.48550/arXiv.2203.07258},
archivePrefix = {arXiv},
       eprint = {2203.07258},
 primaryClass = {astro-ph.CO},
       adsurl = {https://ui.adsabs.harvard.edu/abs/2022arXiv220307258K}
}

@ARTICLE{madau14,
   author = {{Madau}, P. and {Dickinson}, M.},
    title = "{Cosmic Star-Formation History}",
  journal = {\araa},
archivePrefix = "arXiv",
   eprint = {1403.0007},
     year = 2014,
    month = aug,
   volume = 52,
    pages = {415-486},
      doi = {10.1146/annurev-astro-081811-125615},
   adsurl = {http://adsabs.harvard.edu/abs/2014ARA%26A..52..415M}
}

@inproceedings{piperOverview,
  title={The primordial inflation polarization explorer (PIPER)},
  author={Gandilo, Natalie N and Ade, Peter AR and Benford, Dominic and Bennett, Charles L and Chuss, David T and Dotson, Jessie L and Eimer, Joseph R and Fixsen, Dale J and Halpern, Mark and Hilton, Gene and others},
  booktitle={Millimeter, Submillimeter, and Far-Infrared Detectors and Instrumentation for Astronomy VIII},
  volume={9914},
  pages={372--379},
  year={2016},
  organization={SPIE}
}

@ARTICLE{zavala+21,
       author = {{Zavala}, J.~A. and {Casey}, C.~M. and {Manning}, S.~M. and {Aravena}, M. and {Bethermin}, M. and {Caputi}, K.~I. and {Clements}, D.~L. and {Cunha}, E. da and {Drew}, P. and {Finkelstein}, S.~L. and {Fujimoto}, S. and {Hayward}, C. and {Hodge}, J. and {Kartaltepe}, J.~S. and {Knudsen}, K. and {Koekemoer}, A.~M. and {Long}, A.~S. and {Magdis}, G.~E. and {Man}, A.~W.~S. and {Popping}, G. and {Sanders}, D. and {Scoville}, N. and {Sheth}, K. and {Staguhn}, J. and {Toft}, S. and {Treister}, E. and {Vieira}, J.~D. and {Yun}, M.~S.},
        title = "{The Evolution of the IR Luminosity Function and Dust-obscured Star Formation over the Past 13 Billion Years}",
      journal = {\apj},
         year = 2021,
        month = mar,
       volume = {909},
       number = {2},
          eid = {165},
        pages = {165},
          doi = {10.3847/1538-4357/abdb27},
archivePrefix = {arXiv},
       eprint = {2101.04734},
 primaryClass = {astro-ph.GA},
       adsurl = {https://ui.adsabs.harvard.edu/abs/2021ApJ...909..165Z}
}

@ARTICLE{gruppioni+20,
       author = {{Gruppioni}, C. and {B{\'e}thermin}, M. and {Loiacono}, F. and {Le F{\`e}vre}, O. and {Capak}, P. and {Cassata}, P. and {Faisst}, A.~L. and {Schaerer}, D. and {Silverman}, J. and {Yan}, L. and {Bardelli}, S. and {Boquien}, M. and {Carraro}, R. and {Cimatti}, A. and {Dessauges-Zavadsky}, M. and {Ginolfi}, M. and {Fujimoto}, S. and {Hathi}, N.~P. and {Jones}, G.~C. and {Khusanova}, Y. and {Koekemoer}, A.~M. and {Lagache}, G. and {Lemaux}, B.~C. and {Oesch}, P.~A. and {Pozzi}, F. and {Riechers}, D.~A. and {Rodighiero}, G. and {Romano}, M. and {Talia}, M. and {Vallini}, L. and {Vergani}, D. and {Zamorani}, G. and {Zucca}, E.},
        title = "{The ALPINE-ALMA [CII] survey. The nature, luminosity function, and star formation history of dusty galaxies up to z ~ 6}",
      journal = {\aap},
         year = 2020,
        month = nov,
       volume = {643},
          eid = {A8},
        pages = {A8},
          doi = {10.1051/0004-6361/202038487},
archivePrefix = {arXiv},
       eprint = {2006.04974},
 primaryClass = {astro-ph.GA},
       adsurl = {https://ui.adsabs.harvard.edu/abs/2020A&A...643A...8G}
}

@ARTICLE{walter12,
   author = {{Walter}, F. and {Decarli}, R. and {Carilli}, C. and {Bertoldi}, F. and {Cox}, P. and {da Cunha}, E. and {Daddi}, E. and {Dickinson}, M. and {Downes}, D. and {Elbaz}, D. and {Ellis}, R. and {Hodge}, J. and {Neri}, R. and {Riechers}, D.~A. and {Weiss}, A. and {Bell}, E. and {Dannerbauer}, H. and {Krips}, M. and {Krumholz}, M. and {Lentati}, L. and {Maiolino}, R. and {Menten}, K. and {Rix}, H.-W. and {Robertson}, B. and {Spinrad}, H. and {Stark}, D.~P. and {Stern}, D.},
    title = "{The intense starburst HDF850.1 in a galaxy overdensity at $z \approx 5.2$ in the Hubble Deep Field}",
  journal = {\nat},
archivePrefix = "arXiv",
   eprint = {1206.2641},
     year = 2012,
    month = jun,
   volume = 486,
    pages = {233-236},
      doi = {10.1038/nature11073},
   adsurl = {http://adsabs.harvard.edu/abs/2012Natur.486..233W}
}

@ARTICLE{marrone18,
   author = {{Marrone}, D.~P. and {Spilker}, J.~S. and {Hayward}, C.~C. and 
	{Vieira}, J.~D. and {Aravena}, M. and {Ashby}, M.~L.~N. and 
	{Bayliss}, M.~B. and {B{\'e}thermin}, M. and {Brodwin}, M. and 
	{Bothwell}, M.~S. and {Carlstrom}, J.~E. and {Chapman}, S.~C. and 
	{Chen}, C.-C. and {Crawford}, T.~M. and {Cunningham}, D.~J.~M. and 
	{De Breuck}, C. and {Fassnacht}, C.~D. and {Gonzalez}, A.~H. and 
	{Greve}, T.~R. and {Hezaveh}, Y.~D. and {Lacaille}, K. and {Litke}, K.~C. and 
	{Lower}, S. and {Ma}, J. and {Malkan}, M. and {Miller}, T.~B. and 
	{Morningstar}, W.~R. and {Murphy}, E.~J. and {Narayanan}, D. and 
	{Phadke}, K.~A. and {Rotermund}, K.~M. and {Sreevani}, J. and 
	{Stalder}, B. and {Stark}, A.~A. and {Strandet}, M.~L. and {Tang}, M. and 
	{Wei{\ss}}, A.},
    title = "{Galaxy growth in a massive halo in the first billion years of cosmic history}",
  journal = {\nat},
archivePrefix = "arXiv",
   eprint = {1712.03020},
     year = 2018,
    month = jan,
   volume = 553,
    pages = {51-54},
      doi = {10.1038/nature24629},
   adsurl = {http://adsabs.harvard.edu/abs/2018Natur.553...51M}
}

@article{masui2013,
  title={Measurement of 21 cm brightness fluctuations at z~ 0.8 in cross-correlation},
  author={Masui, KW and Switzer, ER and Banavar, N and Bandura, K and Blake, C and Calin, L-M and Chang, T-C and Chen, X and Li, Y-C and Liao, Y-W and others},
  journal={The Astrophysical Journal Letters},
  volume={763},
  number={1},
  pages={L20},
  year={2013},
  publisher={IOP Publishing}
}

@ARTICLE{mowla3dDASH,
       author = {{Mowla}, Lamiya A. and {Cutler}, Sam E. and {Brammer}, Gabriel B. and {Momcheva}, Ivelina G. and {Whitaker}, Katherine E. and {van Dokkum}, Pieter G. and {Bezanson}, Rachel S. and {F{\"o}rster Schreiber}, Natascha M. and {Franx}, Marijn and {Iyer}, Kartheik G. and {Marchesini}, Danilo and {Muzzin}, Adam and {Nelson}, Erica J. and {Skelton}, Rosalind E. and {Snyder}, Gregory F. and {Wake}, David A. and {Wuyts}, Stijn and {van der Wel}, Arjen},
        title = "{3D-DASH: The Widest Near-infrared Hubble Space Telescope Survey}",
      journal = {\apj},
         year = 2022,
        month = jul,
       volume = {933},
       number = {2},
          eid = {129},
        pages = {129},
          doi = {10.3847/1538-4357/ac71af},
archivePrefix = {arXiv},
       eprint = {2206.01156},
 primaryClass = {astro-ph.GA},
       adsurl = {https://ui.adsabs.harvard.edu/abs/2022ApJ...933..129M}
}

@ARTICLE{williams+23,
       author = {{Williams}, Christina C. and {Alberts}, Stacey and {Ji}, Zhiyuan and {Hainline}, Kevin N. and {Lyu}, Jianwei and {Rieke}, George and {Endsley}, Ryan and {Suess}, Katherine A. and {Johnson}, Benjamin D. and {Florian}, Michael and {Shivaei}, Irene and {Rujopakarn}, Wiphu and {Baker}, William M. and {Bhatawdekar}, Rachana and {Boyett}, Kristan and {Bunker}, Andrew J. and {Carniani}, Stefano and {Charlot}, Stephane and {Curtis-Lake}, Emma and {DeCoursey}, Christa and {de Graaff}, Anna and {Egami}, Eiichi and {Eisenstein}, Daniel J. and {Gibson}, Justus L. and {Hausen}, Ryan and {Helton}, Jakob M. and {Maiolino}, Roberto and {Maseda}, Michael V. and {Nelson}, Erica J. and {Perez-Gonzalez}, Pablo G. and {Rieke}, Marcia J. and {Robertson}, Brant E. and {Sun}, Fengwu and {Tacchella}, Sandro and {Willmer}, Christopher N.~A. and {Willott}, Chris J.},
        title = "{The galaxies missed by Hubble and ALMA: the contribution of extremely red galaxies to the cosmic census at 3<z<8}",
      journal = {arXiv e-prints},
         year = 2023,
        month = nov,
          eid = {arXiv:2311.07483},
        pages = {arXiv:2311.07483},
          doi = {10.48550/arXiv.2311.07483},
archivePrefix = {arXiv},
       eprint = {2311.07483},
 primaryClass = {astro-ph.GA},
       adsurl = {https://ui.adsabs.harvard.edu/abs/2023arXiv231107483W}
}

@ARTICLE{harikane+23,
       author = {{Harikane}, Yuichi and {Ouchi}, Masami and {Oguri}, Masamune and {Ono}, Yoshiaki and {Nakajima}, Kimihiko and {Isobe}, Yuki and {Umeda}, Hiroya and {Mawatari}, Ken and {Zhang}, Yechi},
        title = "{A Comprehensive Study of Galaxies at z   9-16 Found in the Early JWST Data: Ultraviolet Luminosity Functions and Cosmic Star Formation History at the Pre-reionization Epoch}",
      journal = {\apjs},
         year = 2023,
        month = mar,
       volume = {265},
       number = {1},
          eid = {5},
        pages = {5},
          doi = {10.3847/1538-4365/acaaa9},
archivePrefix = {arXiv},
       eprint = {2208.01612},
 primaryClass = {astro-ph.GA},
       adsurl = {https://ui.adsabs.harvard.edu/abs/2023ApJS..265....5H}
}

@ARTICLE{loiacono+21,
       author = {{Loiacono}, Federica and {Decarli}, Roberto and {Gruppioni}, Carlotta and {Talia}, Margherita and {Cimatti}, Andrea and {Zamorani}, Gianni and {Pozzi}, Francesca and {Yan}, Lin and {Lemaux}, Brian C. and {Riechers}, Dominik A. and {Le F{\`e}vre}, Olivier and {B{\`e}thermin}, Matthieu and {Capak}, Peter and {Cassata}, Paolo and {Faisst}, Andreas and {Schaerer}, Daniel and {Silverman}, John D. and {Bardelli}, Sandro and {Boquien}, M{\'e}d{\'e}ric and {Burkutean}, Sandra and {Dessauges-Zavadsky}, Miroslava and {Fudamoto}, Yoshinobu and {Fujimoto}, Seiji and {Ginolfi}, Michele and {Hathi}, Nimish P. and {Jones}, Gareth C. and {Khusanova}, Yana and {Koekemoer}, Anton M. and {Lagache}, Guilaine and {Lubin}, Lori M. and {Massardi}, Marcella and {Oesch}, Pascal and {Romano}, Michael and {Vallini}, Livia and {Vergani}, Daniela and {Zucca}, Elena},
        title = "{The ALPINE-ALMA [C II] survey. Luminosity function of serendipitous [C II] line emitters at z {\ensuremath{\sim}} 5}",
      journal = {\aap},
         year = 2021,
        month = feb,
       volume = {646},
          eid = {A76},
        pages = {A76},
          doi = {10.1051/0004-6361/202038607},
archivePrefix = {arXiv},
       eprint = {2006.04837},
 primaryClass = {astro-ph.GA},
       adsurl = {https://ui.adsabs.harvard.edu/abs/2021A&A...646A..76L}
}

@ARTICLE{spilker22,
       author = {{Spilker}, Justin S. and {Hayward}, Christopher C. and {Marrone}, Daniel P. and {Aravena}, Manuel and {B{\'e}thermin}, Matthieu and {Burgoyne}, James and {Chapman}, Scott C. and {Greve}, Thomas R. and {Gururajan}, Gayathri and {Hezaveh}, Yashar D. and {Hill}, Ryley and {Litke}, Katrina C. and {Lovell}, Christopher C. and {Malkan}, Matthew A. and {Murphy}, Eric J. and {Narayanan}, Desika and {Phadke}, Kedar A. and {Reuter}, Cassie and {Stark}, Antony A. and {Sulzenauer}, Nikolaus and {Vieira}, Joaquin D. and {Vizgan}, David and {Wei{\ss}}, Axel},
        title = "{Chaotic and Clumpy Galaxy Formation in an Extremely Massive Reionization-era Halo}",
      journal = {\apjl},
         year = 2022,
        month = apr,
       volume = {929},
       number = {1},
          eid = {L3},
        pages = {L3},
          doi = {10.3847/2041-8213/ac61e6},
archivePrefix = {arXiv},
       eprint = {2203.14972},
 primaryClass = {astro-ph.GA},
       adsurl = {https://ui.adsabs.harvard.edu/abs/2022ApJ...929L...3S}
}

@ARTICLE{lefevre+20,
       author = {{Le F{\`e}vre}, O. and {B{\'e}thermin}, M. and {Faisst}, A. and {Jones}, G.~C. and {Capak}, P. and {Cassata}, P. and {Silverman}, J.~D. and {Schaerer}, D. and {Yan}, L. and {Amorin}, R. and {Bardelli}, S. and {Boquien}, M. and {Cimatti}, A. and {Dessauges-Zavadsky}, M. and {Giavalisco}, M. and {Hathi}, N.~P. and {Fudamoto}, Y. and {Fujimoto}, S. and {Ginolfi}, M. and {Gruppioni}, C. and {Hemmati}, S. and {Ibar}, E. and {Koekemoer}, A. and {Khusanova}, Y. and {Lagache}, G. and {Lemaux}, B.~C. and {Loiacono}, F. and {Maiolino}, R. and {Mancini}, C. and {Narayanan}, D. and {Morselli}, L. and {M{\'e}ndez-Hern{\`a}ndez}, Hugo and {Oesch}, P.~A. and {Pozzi}, F. and {Romano}, M. and {Riechers}, D. and {Scoville}, N. and {Talia}, M. and {Tasca}, L.~A.~M. and {Thomas}, R. and {Toft}, S. and {Vallini}, L. and {Vergani}, D. and {Walter}, F. and {Zamorani}, G. and {Zucca}, E.},
        title = "{The ALPINE-ALMA [CII] survey. Survey strategy, observations, and sample properties of 118 star-forming galaxies at 4 < z < 6}",
      journal = {\aap},
         year = 2020,
        month = nov,
       volume = {643},
          eid = {A1},
        pages = {A1},
          doi = {10.1051/0004-6361/201936965},
archivePrefix = {arXiv},
       eprint = {1910.09517},
 primaryClass = {astro-ph.CO},
       adsurl = {https://ui.adsabs.harvard.edu/abs/2020A&A...643A...1L}
}

@article{Hemmati17,
doi = {10.3847/1538-4357/834/1/36},
url = {https://doi.org/10.3847/1538-4357/834/1/36},
year = {2017},
month = {jan},
publisher = {The American Astronomical Society},
volume = {834},
number = {1},
pages = {36},
author = {Hemmati, Shoubaneh and Yan, Lin and Diaz-Santos, Tanio and Armus, Lee and Capak, Peter and Faisst, Andreas and Masters, Daniel},
title = {THE LOCAL [C ii] 158 μm EMISSION LINE LUMINOSITY FUNCTION},
journal = {The Astrophysical Journal}
}

@ARTICLE{herrera-camus+15,
       author = {{Herrera-Camus}, R. and {Bolatto}, A.~D. and {Wolfire}, M.~G. and {Smith}, J.~D. and {Croxall}, K.~V. and {Kennicutt}, R.~C. and {Calzetti}, D. and {Helou}, G. and {Walter}, F. and {Leroy}, A.~K. and {Draine}, B. and {Brandl}, B.~R. and {Armus}, L. and {Sandstrom}, K.~M. and {Dale}, D.~A. and {Aniano}, G. and {Meidt}, S.~E. and {Boquien}, M. and {Hunt}, L.~K. and {Galametz}, M. and {Tabatabaei}, F.~S. and {Murphy}, E.~J. and {Appleton}, P. and {Roussel}, H. and {Engelbracht}, C. and {Beirao}, P.},
        title = "{[C II] 158 {\ensuremath{\mu}}m Emission as a Star Formation Tracer}",
      journal = {\apj},
         year = 2015,
        month = feb,
       volume = {800},
       number = {1},
          eid = {1},
        pages = {1},
          doi = {10.1088/0004-637X/800/1/1},
archivePrefix = {arXiv},
       eprint = {1409.7123},
 primaryClass = {astro-ph.GA},
       adsurl = {https://ui.adsabs.harvard.edu/abs/2015ApJ...800....1H}
}

@ARTICLE{delooze+14,
       author = {{De Looze}, Ilse and {Cormier}, Diane and {Lebouteiller}, Vianney and {Madden}, Suzanne and {Baes}, Maarten and {Bendo}, George J. and {Boquien}, M{\'e}d{\'e}ric and {Boselli}, Alessandro and {Clements}, David L. and {Cortese}, Luca and {Cooray}, Asantha and {Galametz}, Maud and {Galliano}, Fr{\'e}d{\'e}ric and {Graci{\'a}-Carpio}, Javier and {Isaak}, Kate and {Karczewski}, Oskar {\L}. and {Parkin}, Tara J. and {Pellegrini}, Eric W. and {R{\'e}my-Ruyer}, Aur{\'e}lie and {Spinoglio}, Luigi and {Smith}, Matthew W.~L. and {Sturm}, Eckhard},
        title = "{The applicability of far-infrared fine-structure lines as star formation rate tracers over wide ranges of metallicities and galaxy types}",
      journal = {\aap},
         year = 2014,
        month = aug,
       volume = {568},
          eid = {A62},
        pages = {A62},
          doi = {10.1051/0004-6361/201322489},
archivePrefix = {arXiv},
       eprint = {1402.4075},
 primaryClass = {astro-ph.GA},
       adsurl = {https://ui.adsabs.harvard.edu/abs/2014A&A...568A..62D}
}

@article{Kennicutt12,
   title={Star Formation in the Milky Way and Nearby Galaxies},
   volume={50},
   ISSN={1545-4282},
   url={http://dx.doi.org/10.1146/annurev-astro-081811-125610},
   DOI={10.1146/annurev-astro-081811-125610},
   number={1},
   journal={Annual Review of Astronomy and Astrophysics},
   publisher={Annual Reviews},
   author={Kennicutt, Robert C. and Evans, Neal J.},
   year={2012},
   month=sep, pages={531–608} }

@ARTICLE{pullen18,
       author = {{Pullen}, Anthony R. and {Serra}, Paolo and {Chang}, Tzu-Ching and {Dor{\'e}}, Olivier and {Ho}, Shirley},
        title = "{Search for C II emission on cosmological scales at redshift Z {\ensuremath{\sim}} 2.6}",
      journal = {\mnras},
         year = 2018,
        month = aug,
       volume = {478},
       number = {2},
        pages = {1911-1924},
          doi = {10.1093/mnras/sty1243},
archivePrefix = {arXiv},
       eprint = {1707.06172},
 primaryClass = {astro-ph.CO},
       adsurl = {https://ui.adsabs.harvard.edu/abs/2018MNRAS.478.1911P}
}

@article{Pullen13,
doi = {10.1088/0004-637X/768/1/15},
url = {https://dx.doi.org/10.1088/0004-637X/768/1/15},
year = {2013},
month = {apr},
publisher = {The American Astronomical Society},
volume = {768},
number = {1},
pages = {15},
author = {Pullen, Anthony R. and Chang, Tzu-Ching and Doré, Olivier and Lidz, Adam},
title = {CROSS-CORRELATIONS AS A COSMOLOGICAL CARBON MONOXIDE DETECTOR},
journal = {The Astrophysical Journal}
}

@ARTICLE{PRIMA_concept,
       author = {{Glenn}, Jason and {Meixner}, Margaret and {Bradford}, Charles M. and {Pontoppidan}, Klaus and {Pope}, Alexandra and {Kataria}, Tiffany and {Rocca}, Jennifer and {Luthman}, Elizabeth and {Armus}, Lee and {Baselmans}, Jochem and {Battersby}, Cara and {Bollato}, Alberto and {Burgarella}, Denis and {Chen}, Weibo and {Ciesla}, Laure and {Day}, Peter and {Di Giorgio}, Anna and {Dipirro}, Michael and {Dowell}, Charles Darren and {Echternach}, Pierre and {Essinger-Hileman}, Thomas and {Foote}, Marc and {Gruppioni}, Carlotta and {Hensley}, Brandon and {Henning}, Thomas and {Jellema}, Willem and {Johnson}, Matthew and {Kogut}, Alan and {Krause}, Oliver and {McGuire}, James and {Mills}, Elisabeth and {Moullet}, Arielle and {Rodgers}, Michael and {Sauvage}, Marc and {Smith}, John D. and {Somerville}, Rachel and {Staguhn}, Johannes and {Stevenson}, Thomas and {Tucker}, Carole and {Unwin}, Stephen and {Ziemer}, John and {Cannella}, Matthew and {Dissly}, Richard},
        title = "{PRIMA mission concept}",
      journal = {Journal of Astronomical Telescopes, Instruments, and Systems},
         year = 2025,
        month = jul,
       volume = {11},
          eid = {031628},
        pages = {031628},
          doi = {10.1117/1.JATIS.11.3.031628},
       adsurl = {https://ui.adsabs.harvard.edu/abs/2025JATIS..11c1628G}
}

@article{Merlin_2021,
   title={The ASTRODEEP-GS43 catalogue: New photometry and redshifts for the CANDELS GOODS-South field},
   volume={649},
   ISSN={1432-0746},
   url={http://dx.doi.org/10.1051/0004-6361/202140310},
   DOI={10.1051/0004-6361/202140310},
   journal={\aap},
   publisher={EDP Sciences},
   author={Merlin, E. and Castellano, M. and Santini, P. and Cipolletta, G. and Boutsia, K. and Schreiber, C. and Buitrago, F. and Fontana, A. and Elbaz, D. and Dunlop, J. and Grazian, A. and McLure, R. and McLeod, D. and Nonino, M. and Milvang-Jensen, B. and Derriere, S. and Hathi, N. P. and Pentericci, L. and Fortuni, F. and Calabrò, A.},
   year={2021},
   month=may, pages={A22} }

@ARTICLE{MD_2014,
       author = {{Madau}, Piero and {Dickinson}, Mark},
        title = "{Cosmic Star-Formation History}",
      journal = {\araa},
         year = 2014,
        month = aug,
       volume = {52},
        pages = {415-486},
          doi = {10.1146/annurev-astro-081811-125615},
archivePrefix = {arXiv},
       eprint = {1403.0007},
 primaryClass = {astro-ph.CO},
       adsurl = {https://ui.adsabs.harvard.edu/abs/2014ARA&A..52..415M}
}

@software{CAMBL,
       author = {{Lewis}, Antony and {Challinor}, Anthony},
        title = "{CAMB: Code for Anisotropies in the Microwave Background}",
 howpublished = {Astrophysics Source Code Library, record ascl:1102.026},
         year = 2011,
        month = feb,
          eid = {ascl:1102.026},
       adsurl = {https://ui.adsabs.harvard.edu/abs/2011ascl.soft02026L}
}

@article{De_Looze_2014,
   title={The applicability of far-infrared fine-structure lines as star formation rate tracers over wide ranges of metallicities and galaxy types},
   volume={568},
   ISSN={1432-0746},
   url={http://dx.doi.org/10.1051/0004-6361/201322489},
   DOI={10.1051/0004-6361/201322489},
   journal={\aap},
   publisher={EDP Sciences},
   author={De Looze, Ilse and Cormier, Diane and Lebouteiller, Vianney and Madden, Suzanne and Baes, Maarten and Bendo, George J. and Boquien, Médéric and Boselli, Alessandro and Clements, David L. and Cortese, Luca and Cooray, Asantha and Galametz, Maud and Galliano, Frédéric and Graciá-Carpio, Javier and Isaak, Kate and Karczewski, Oskar Ł. and Parkin, Tara J. and Pellegrini, Eric W. and Rémy-Ruyer, Aurélie and Spinoglio, Luigi and Smith, Matthew W. L. and Sturm, Eckhard},
   year={2014},
   month=aug, pages={A62} }

@article{Jullo12,
doi = {10.1088/0004-637X/750/1/37},
url = {https://dx.doi.org/10.1088/0004-637X/750/1/37},
year = {2012},
month = {apr},
publisher = {The American Astronomical Society},
volume = {750},
number = {1},
pages = {37},
author = {Jullo, Eric and Rhodes, Jason and Kiessling, Alina and Taylor, James E. and Massey, Richard and Berge, Joel and Schimd, Carlo and Kneib, Jean-Paul and Scoville, Nick},
title = {COSMOS: STOCHASTIC BIAS FROM MEASUREMENTS OF WEAK LENSING AND GALAXY CLUSTERING},
journal = {The Astrophysical Journal}
}

@article{Planck18,
   title={Planck2018 results: VI. Cosmological parameters},
   volume={641},
   ISSN={1432-0746},
   url={http://dx.doi.org/10.1051/0004-6361/201833910},
   DOI={10.1051/0004-6361/201833910},
   journal={\aap},
   publisher={EDP Sciences},
   author={Aghanim, N. and Akrami, Y. and Ashdown, M. and Aumont, J. and Baccigalupi, C. and Ballardini, M. and Banday, A. J. and Barreiro, R. B. and Bartolo, N. and Basak, S. and Battye, R. and Benabed, K. and Bernard, J.-P. and Bersanelli, M. and Bielewicz, P. and Bock, J. J. and Bond, J. R. and Borrill, J. and Bouchet, F. R. and Boulanger, F. and Bucher, M. and Burigana, C. and Butler, R. C. and Calabrese, E. and Cardoso, J.-F. and Carron, J. and Challinor, A. and Chiang, H. C. and Chluba, J. and Colombo, L. P. L. and Combet, C. and Contreras, D. and Crill, B. P. and Cuttaia, F. and de Bernardis, P. and de Zotti, G. and Delabrouille, J. and Delouis, J.-M. and Di Valentino, E. and Diego, J. M. and Doré, O. and Douspis, M. and Ducout, A. and Dupac, X. and Dusini, S. and Efstathiou, G. and Elsner, F. and Enßlin, T. A. and Eriksen, H. K. and Fantaye, Y. and Farhang, M. and Fergusson, J. and Fernandez-Cobos, R. and Finelli, F. and Forastieri, F. and Frailis, M. and Fraisse, A. A. and Franceschi, E. and Frolov, A. and Galeotta, S. and Galli, S. and Ganga, K. and Génova-Santos, R. T. and Gerbino, M. and Ghosh, T. and González-Nuevo, J. and Górski, K. M. and Gratton, S. and Gruppuso, A. and Gudmundsson, J. E. and Hamann, J. and Handley, W. and Hansen, F. K. and Herranz, D. and Hildebrandt, S. R. and Hivon, E. and Huang, Z. and Jaffe, A. H. and Jones, W. C. and Karakci, A. and Keihänen, E. and Keskitalo, R. and Kiiveri, K. and Kim, J. and Kisner, T. S. and Knox, L. and Krachmalnicoff, N. and Kunz, M. and Kurki-Suonio, H. and Lagache, G. and Lamarre, J.-M. and Lasenby, A. and Lattanzi, M. and Lawrence, C. R. and Le Jeune, M. and Lemos, P. and Lesgourgues, J. and Levrier, F. and Lewis, A. and Liguori, M. and Lilje, P. B. and Lilley, M. and Lindholm, V. and López-Caniego, M. and Lubin, P. M. and Ma, Y.-Z. and Macías-Pérez, J. F. and Maggio, G. and Maino, D. and Mandolesi, N. and Mangilli, A. and Marcos-Caballero, A. and Maris, M. and Martin, P. G. and Martinelli, M. and Martínez-González, E. and Matarrese, S. and Mauri, N. and McEwen, J. D. and Meinhold, P. R. and Melchiorri, A. and Mennella, A. and Migliaccio, M. and Millea, M. and Mitra, S. and Miville-Deschênes, M.-A. and Molinari, D. and Montier, L. and Morgante, G. and Moss, A. and Natoli, P. and Nørgaard-Nielsen, H. U. and Pagano, L. and Paoletti, D. and Partridge, B. and Patanchon, G. and Peiris, H. V. and Perrotta, F. and Pettorino, V. and Piacentini, F. and Polastri, L. and Polenta, G. and Puget, J.-L. and Rachen, J. P. and Reinecke, M. and Remazeilles, M. and Renzi, A. and Rocha, G. and Rosset, C. and Roudier, G. and Rubiño-Martín, J. A. and Ruiz-Granados, B. and Salvati, L. and Sandri, M. and Savelainen, M. and Scott, D. and Shellard, E. P. S. and Sirignano, C. and Sirri, G. and Spencer, L. D. and Sunyaev, R. and Suur-Uski, A.-S. and Tauber, J. A. and Tavagnacco, D. and Tenti, M. and Toffolatti, L. and Tomasi, M. and Trombetti, T. and Valenziano, L. and Valiviita, J. and Van Tent, B. and Vibert, L. and Vielva, P. and Villa, F. and Vittorio, N. and Wandelt, B. D. and Wehus, I. K. and White, M. and White, S. D. M. and Zacchei, A. and Zonca, A.},
   year={2020},
   month=sep, pages={A6} }

@article{keenan20,
  title={Biases and cosmic variance in molecular gas abundance measurements at high redshift},
  author={Keenan, Ryan P and Marrone, Daniel P and Keating, Garrett K},
  journal={The Astrophysical Journal},
  volume={904},
  number={2},
  pages={127},
  year={2020},
  publisher={IOP Publishing}
}

@article{keenan22,
  title={An Intensity Mapping Constraint on the CO-galaxy Cross-power Spectrum at Redshift~ 3},
  author={Keenan, Ryan P and Keating, Garrett K and Marrone, Daniel P},
  journal={The Astrophysical Journal},
  volume={927},
  number={2},
  pages={161},
  year={2022},
  publisher={IOP Publishing}
}

@ARTICLE{Kamenetzky16,
       author = {{Kamenetzky}, J. and {Rangwala}, N. and {Glenn}, J. and {Maloney}, P.~R. and {Conley}, A.},
        title = "{L‧$_{CO}$/L$_{FIR}$ Relations with CO Rotational Ladders of Galaxies Across the Herschel SPIRE Archive}",
      journal = {\apj},
         year = 2016,
        month = oct,
       volume = {829},
       number = {2},
          eid = {93},
        pages = {93},
          doi = {10.3847/0004-637X/829/2/93},
archivePrefix = {arXiv},
       eprint = {1508.05102},
 primaryClass = {astro-ph.GA},
       adsurl = {https://ui.adsabs.harvard.edu/abs/2016ApJ...829...93K}
}

@ARTICLE{keating16,
       author = {{Keating}, Garrett K. and {Marrone}, Daniel P. and {Bower}, Geoffrey C. and {Leitch}, Erik and {Carlstrom}, John E. and {DeBoer}, David R.},
        title = "{COPSS II: The Molecular Gas Content of Ten Million Cubic Megaparsecs at Redshift z {\ensuremath{\sim}} 3}",
      journal = {\apj},
         year = 2016,
        month = oct,
       volume = {830},
       number = {1},
          eid = {34},
        pages = {34},
          doi = {10.3847/0004-637X/830/1/34},
archivePrefix = {arXiv},
       eprint = {1605.03971},
 primaryClass = {astro-ph.GA},
       adsurl = {https://ui.adsabs.harvard.edu/abs/2016ApJ...830...34K}
}

@article{keating20,
	title = {An {Intensity} {Mapping} {Detection} of {Aggregate} {CO} {Line} {Emission} at 3 mm},
	volume = {901},
	url = {https://dx.doi.org/10.3847/1538-4357/abb08e},
	doi = {10.3847/1538-4357/abb08e},
	number = {2},
	journal = {The Astrophysical Journal},
	author = {Keating, Garrett K. and Marrone, Daniel P. and Bower, Geoffrey C. and Keenan, Ryan P.},
	month = oct,
	year = {2020},
	note = {Publisher: The American Astronomical Society},
	pages = {141},
}

@article{vieira20,
  title={The Terahertz Intensity Mapper (TIM): A next-generation experiment for galaxy evolution studies},
  author={Vieira, Joaquin and Aguirre, James and Bradford, C Matt and Filippini, Jeffrey and Groppi, Christopher and Marrone, Dan and Bethermin, Matthieu and Chang, Tzu-Ching and Devlin, Mark and Dore, Oliver and others},
  journal={arXiv preprint arXiv:2009.14340},
  year={2020}
}

@inproceedings{marrone2022terahertz,
  title={The terahertz intensity mapper: a balloon-borne imaging spectrometer for galaxy evolution},
  author={Marrone, Daniel P and Aguirre, James E and Bracks, Justin S and Bradford, Charles M and Brendal, Brockton S and Bumble, Bruce and Corso, Anthony J and Devlin, Mark J and Emerson, Nick and Filippini, Jeffrey P and others},
  booktitle={Millimeter, Submillimeter, and Far-Infrared Detectors and Instrumentation for Astronomy XI},
  volume={12190},
  pages={131--142},
  year={2022},
  organization={SPIE}
}
\bibliographystyle{aasjournal}




\end{document}